\documentclass[a4paper,fleqn,usenatbib]{mnras}

\usepackage[T1]{fontenc}
\usepackage{ae,aecompl}

\usepackage{graphicx}	
\usepackage{amsmath}	
\usepackage{amssymb}	

\usepackage{txfonts}
\usepackage{pifont}
\usepackage{gensymb}
\usepackage{color}

\newcommand{\cmark}{\ding{51}}%
\newcommand{\xmark}{\ding{55}}%

\title[A Transient Source in Leoncino Dwarf]{An Unusual Transient in the Extremely Metal-Poor Galaxy SDSS~J094332.35+332657.6 (Leoncino Dwarf)}

\author[Filho \& S\'anchez Almeida]{
Mercedes E. Filho,$^{1,2}$\thanks{E-mail: mfilho@fe.up.pt (MEF)}
and J. S\'anchez Almeida,$^{3,4}$
\\
$^{1}$Center for Astrophysics and Gravitation - CENTRA/SIM, Departamento de F\'\i sica, Instituto Superior T\'ecnico, Universidade de Lisboa, Av. Rovisco Pais 1, \\
P-1049-001 Lisbon, Portugal \\
$^{2}$Departamento de Engenharia F\'\i sica, Universidade do Porto, Faculdade de Engenharia, Universidade do Porto, Rua Dr. Roberto Frias, s/n, P-4200-465, \\
Oporto, Portugal\\
$^{3}$Instituto Astrof\'\i sica de Canarias, 38205 La Laguna, Tenerife, Spain\\
$^{4}$Departamento de Astrof\'\i sica, Universidad La Laguna, 38206 La Laguna, Tenerife, Spain
}

\date{Accepted XXX. Received YYY; in original form ZZZ}

\pubyear{2017}

\begin{document}
\label{firstpage}
\pagerange{\pageref{firstpage}--\pageref{lastpage}}
\maketitle


\begin{abstract}

We have serendipitously discovered that Leoncino Dwarf, an ultra-faint, low-metallicity record-holder dwarf galaxy, may have hosted a transient source, and possibly exhibited a change in morphology, a shift in the center of brightness, and peak variability of the main (host) source in images taken approximately 40 yr apart; it is highly likely that these phenomena are related. Scenarios involving a Solar System object, a stellar cluster, dust enshrouding, and accretion variability have been considered, and discarded, as the origin of the transient. Although a combination of time-varying strong and weak lensing effects, induced by an intermediate mass black hole (10$^4$ -- 5 $\times$ 10$^{5}$ M$_{\odot}$) moving within the Milky Way halo (0.1 -- 4 kpc), can conceivably explain all of the observed variable galaxy properties, it is statistically highly unlikely according to current theoretical predictions, and, therefore, also discarded. A cataclysmic event such as a supernova/hypernova could have occurred, as long as the event was observed towards the later/late-stage descent of the light curve, but this scenario fails to explain the absence of a post-explosion source and/or host HII region in recent optical images. An episode related to the giant eruption of a luminous blue variable star, a stellar merger or a nova, observed at, or near, peak magnitude may explain the transient source and possibly the change in morphology/center of brightness, but can not justify the main source peak variability, unless stellar variability is evoked.

\end{abstract}


\section{Introduction}

Time-domain astronomy, with an emphasis on transient astronomical events, is one of the major focuses of synoptic, wide-field surveys (e.g., Djorgovski et al. 2012). Transient astronomical events include sources such as SN\footnote{supernovae} of all types, FRBs\footnote{fast radio burst}, GRBs\footnote{gamma-ray burst}, merger events (tidal disruption events, binary system mergers) and novae, amongst others (Sect.~3). Such events are of particular astrophysical interest since they are relatively rare. In some instances, the origin and physical mechanisms that give rise to certain transients is still under debate, and counterparts to many of the sources still remain unknown. Targeted surveys have provided a plethora of new detections, and allowed systematic population studies. In addition, surveys have also discovered new types of transient sources, such as low-luminosity calcium-rich SN (e.g., Perets et al. 2010; Sect.~3), hydrogen-poor superluminous SN (e.g., Quimby et al. 2011), and FRBs (e.g., Lorimer et al. 2007). Recent and current on-going surveys include the Gaia Science Alerts (Hodgkin et al. 2013), the SMT\footnote{SkyMapper Transient} survey (Scalzo et al. 2017), the CRTS\footnote{Catalina Real-Time Transient Survey} (Djorgovski et al. 2011), the CSP\footnote{Carnegie Supernova Project} (Hamuy et al. 2006), the PTF\footnote{Palomar Transient Factory} (Law et al. 2009; Rau et al. 2009) and its sucessor the iPTF\footnote{Intermediate Palomar Transient Factory}, and the Pan-STARRS\footnote{Panoramic Survey Telescope and Rapid Response System} (Chambers et al. 2016). Examples of upcoming projects include the ZTF\footnote{Zwicky Transient Factory}, the CRTS and Pan-STARRS successors, and the LSST\footnote{Large Synoptic Survey Telescope} (Ivezi\'c et al. 2008). Several interesting citizen science projects, included, for example, within Zooniverse\footnote{https://www.zooniverse.org/}, have also provided (or will provide) significant results in the search for transients; these include Astronomy Rewind, Supernova Sighting, Supernova Hunter and Galaxy Zoo Supernova (Smith et al. 2011a). 

This work was motivated by the serendipitous discovery of a Northern knot in the blue POSS\footnote{Palomar Observatory Sky Survey} 1955 image of the extremely metal-poor dwarf galaxy (XMP) SDSS~J094332.35+332657.6 (Fig.~2; left-hand column; row 4 -- 5), a knot which is no longer present in the subsequent POSS (Fig.~2; middle column), HST\footnote{Hubble Space Telescope} (Fig.~3; top) or SDSS\footnote{Sloan Digital Sky Survey} (Fig.~3; bottom) images taken approximately 40 -- 60 years later. SDSS~J094332.35+332657.6 (with the epithet Leoncino Dwarf used hereinafter), the optical counterpart to the HI source AGC~198691, was, until recently (see J0811+4730 with 12+log(O/H) = 6.98$\pm$0.02; Izotov et al. 2017), the local, low-metallicity record-breaker at 12+log(O/H) = 7.02 $\pm$ 0.03 (Hirschauer et al. 2016; hereinafter H16). The XMP category, to which Leoncino Dwarf belongs, is generally composed of low-surface-brightness (S\'anchez Almeida et al. 2017), isolated (Filho et al. 2015) galaxies, currently undergoing star formation processes under extreme low-metallicity conditions (Filho et al. 2013, 2016), likely fueled by cosmological gas accretion (S\'anchez Almeida et a. 2015); hence, XMPs are excellent candidates for the occurrence of extreme and variable events (Sect.~3), some of which may be rare or unusual (e.g., Pustilnik et al. 2008; Izotov \& Thuan 2009; Bomans \& Weis 2011; Pustilnik et al. 2017). Indeed, metal-poor environments appear to be preferred hosts of long GRBs (Fruchter et al. 2006; Modjaz et al. 2008), hydrogen-poor superluminous SN (Lunnan et al. 2014, 2015; Leloudas et al. 2015) and ultraluminous X-ray sources (Sutton et al. 2012). More recently, a recurrent FRB, associated with a persistent radio source of unknown origin, was detected in a metal-poor galaxy (Tendulkar et al. 2017; Bassa et al. 2017; Chatterjee et al. 2017; Marcote et al. 2017; Kokubo et al. 2017). A star-forming metal-poor dwarf galaxy was also found to host the unusual hydrogen-rich SN iPTF14hls (Arcavi et al. 2017); interestingly, the POSS data was used to verify a possible eruption in 1954, besides the 2014 event registered by the iPTF, demonstrating the potential in using long-period-baseline archival data for detecting transient events.

The present paper characterizes the observed properties of Leoncino Dwarf utilizing all the available multi-wavelength data (Sect.~2), and attempts to provide an interpretation for the observed transient phenomenon, as well as the other time-dependent observables (Sect.~3). 

Throughout, a cosmological model where $H_0 = 69.6$ km s$^{-1}$ Mpc$^{-1}$, $\Omega _{\rm M} = 0.286$, and $\Omega _{\rm vac} = 0.714$ (Bennett et al. 2014), has been adopted.


\section{Data}

Motivated by Leoncino Dwarf's identification as one of the most metal-poor dwarf galaxies in the local Universe, multi-wavelength information and data for the target was procured via online database inquiries. A summary of the general properties of Leoncino Dwarf, mainly reproduced from H16, is provided in Table~1. It is to be noted that, although Leoncino Dwarf is too bright and in the large size range, its stellar mass (M$_{\star}$) is within the realm of globular clusters.



\setcounter{table}{0}

\begin{table}

\footnotesize

\begin{center}

\begin{minipage}{87mm}

\caption{General Properties of Leoncino Dwarf}

\begin{tabular}{ l | l || l | l }

\hline

Parameter					& Value											& Parameter					& Value \\

\hline
\hline
		
RA (J2000)					& 09 43 32.4									& $m_{\rm V}$ 				& 19.5 mag \\
Dec (J2000)					& 33 26 58										& $M_{\rm V}$ 				& -10.0 mag \\
12+log(O/H)					& 7.02 $\pm$ 0.03								& $m_{\rm B}$ 				& 19.8	mag \\						
$D^1$						& $\approx$8~Mpc 								& $M_{\rm B}$				& -9.8 mag \\
$d^2$						& 8.1 arcsec									& {\em SB}$_{\rm B}^6$ 		& 24.8 mag arcsec$^{-2}$ \\		
$L({\rm H\alpha}$)$^3$		& 6.4 $\times$ 10$^{37}$ erg s$^{-1}$			& M$_{\star}$				& 1.6 $\times$ 10$^{5}$ M$_{\odot}$ \\
SFR$^4$ 					& 5.1 $\times$ 10$^{-4}$ M$_{\odot}$ yr$^{-1}$	& $V_{\odot}^7$				& 514 $\pm$ 2 km s$^{-1}$	\\	 
$W_{50}^5$					& 33 $\pm$ 2 km s$^{-1}$						& M$_{\rm HI}^8$			& 8.0 $\times$ 10$^{6}$ M$_{\odot}$  \\

\hline

\end{tabular}

$^1$Approximate minimum distance. Leoncino Dwarf is located in the direction of the 'local velocity anomaly', so that its distance is highly uncertain. This distance is used to estimate absolute values in the table. \\
$^2$Angular diameter, taking into account faint extended low surface brightness emission lying beyond the central core of bright stars and the HII region, obtained from the HST images. \\
$^3$H$\alpha$ luminosity obtained from the WIYN narrow-band observations assuming $D \simeq$ 8~Mpc. \\
$^4$Star formation rate computed from the H$\alpha$ luminosity. \\
$^5$ALFALFA HI line width at 50\% of the peak flux density level. \\
$^6$$B$-band surface brightness obtained using the $B$-band apparent magnitude ($m_{\rm B}$) and HST angular diameter ($d$). \\
$^7$ALFALFA HI heliocentric velocity. \\
$^8$ALFALFA HI mass assuming $D \simeq$ 8~Mpc. 

\end{minipage}

\end{center}

\end{table}


\subsection{POSS Data}

There are first (I) and second (II) epoch POSS images of Leoncino Dwarf, taken  approximately 40 years apart (1955 and 1995 -- 1998). These surveys were carried out with the 48-inch (Oschin) Schmidt telescope on Mount Palomar, and registered onto photographic plates, with plate scales of $\sim$1 arcsec pix$^{-1}$. The STScI\footnote{Space Telescope Science Institute} provides photometrically uncalibrated, astrometrically calibrated (through the GSC\footnote{Guide Star Catalog} 2.3; Lasker et al. 2008), digitized POSS images with photometry in the blue (400 -- 500 nm; POSS I-O and POSS II-J), red (600 -- 750 nm; POSS I-E and POSS II-F) and infrared (750 -- 1000 nm; POSS II-N), available via several online catalogs. Table~2 contains the various catalog entries for the POSS photometry of Leoncino Dwarf. 

Astrometrically calibrated, photometrically uncalibrated, 15 arcmin $\times$ 15 arcmin POSS I-O and POSS II-J (blue), POSS I-E and POSS II-F (red), and POSS II-N (infrared) images were uploaded from the STScI, and processed according to the procedure outlined in Section 2.1.1.



\setcounter{table}{1}

\begin{table*}

\footnotesize
\begin{center}

\begin{minipage}{105mm}

\caption{Catalog POSS Photometry for Leoncino Dwarf}

\begin{tabular}{ l c c c c c c c }

\hline

Catalog 	&	Survey		& Band$^{\dagger}$	& $m$ [mag]					& Date 				& Reference \\

\hline
\hline

USNO A2.0	& 	POSS I-O	&	O				& 18.7$\pm$\ldots$^a$ 		& 13/03/1955 		& Monet (1998) \\
USNO B1.0	&	POSS I-O 	&	O				& 19.5$\pm$\ldots$^a$		& 13/03/1955	 	& Monet et al. (2003) \\
USNO B1.0	& 	POSS II-J	&	B$_{\rm J}$		& 19.6$\pm$\ldots$^a$		& 18/03/1996		& Monet et al. (2003) \\
GSC 2.2		& 	POSS II-J	&	B$_{\rm J}$		& 19.0$\pm$0.4				& 18/03/1996		& STScI (2001)$^b$ \\
GSC	2.3.2	&	POSS II-J	& 	B$_{\rm J}$		& 19.0$\pm$0.4				& 18/03/1996 		& STScI (2006)$^b$ \\
GSC	2.3.2	&	POSS II-J	&	B$_{\rm John}$	& 19.5$\pm$0.4				& 18/03/1996 		& STScI (2006)$^b$ \\

\hline 

USNO A2.0	&	POSS I-E	&	E				& 18.9$\pm$\ldots$^a$		& 13/03/1955 		& Monet (1998) \\
USNO B1.0	& 	POSS I-E	&	E				& 19.1$\pm$\ldots$^a$		& 13/03/1955	 	& Monet et al. (2003) \\
USNO B1.0	& 	POSS II-F	&	R$_{\rm F}$		& 19.1$\pm$\ldots$^a$		& 16/04/1998	 	& Monet et al. (2003) \\
GSC 2.3.2	&   POSS II-F	&	R$_{\rm F}$		& 19.5$\pm$0.6		 		& 16/04/1998 		& STScI (2006)$^b$ \\

\hline

USNO B1.0	& 	POSS II-N	&	I$_{\rm N}$		& 18.9$\pm$\ldots$^a$		& 23/02/1995		& Monet et al. (2003) \\

\hline

\end{tabular}

$^{\dagger}$Blue ({\it top}), red ({\it middle}) and infrared ({\it bottom}) bands.\\
$^a$Catalog does not provide a photometric error. \\
$^b$Catalog releases.

\end{minipage}

\end{center}

\end{table*}


\subsubsection{POSS Processing Techniques}

The POSS photometric calibration, as implemented in the USNO\footnote{United States Naval Observatory Astrometric Standards Catalog} A2.0 catalog (Monet 1998), utilizes the Tycho catalog for the bright end of the photometric calibration, while the faint end is set by the USNO CCD\footnote{charge-coupled device} parallax fields in the North and the Yale Southern Proper Motion CCD calibration fields in the South\footnote{http://tdc-www.harvard.edu/catalogs/ua2.html\#phot}. The Tycho-2, GSPC\footnote{Guide Star Photometric Catalog} II and the photometric data measured for the NOFS\footnote{Naval Observatory Flagstaff Station} CCD parallax program are used for the photometric calibration of the USNO B1.0 catalog (Monet et al. 2003), while the Tycho and GSPC (I and II) catalogs are used as photometric calibrators for the GSC 2.3.2 catalog (Lasker et al. 2008). The different photometric calibrations, particularly at the faint end, may explain the $\Delta m \lesssim$ 1~mag in the USNO A2.0 relative to the USNO B1.0/GSC 2.2/GSC 2.3.2 blue photometric values in Table~2. The photometry is optimized for point-like sources; the photometric accuracy is $\approx$0.3 mag for stellar-like objects, with additional errors for faint, extended objects (Monet et al. 2003; Lasker et al. 2008). When compared with the SDSS early data release, photometric offsets for extended, faint objects as large as 0.4 mag were found (Lasker et al. 2008).

The POSS astrometry, as provided by the GSC 2.3 catalog, is tied to the ICRS\footnote{International Celestial Reference System}, defined by the Tycho/Tycho-2 (and a subset of ACT\footnote{Astrographic Catalog/Tycho}) faint-end stars, since the Hipparchus stars are heavily saturated on the plates (Lasker et al. 2008). For this catalog, absolute astrometric errors are typically 0.2 -- 0.3 arcsec for stellar-like sources, with $>$20\% poorer errors for extended, faint sources (Lasker et al. 2008). An initial comparison with the SDSS early data release and UCAC\footnote{United States Naval Observatory Charge-Coupled Device Astrograph Catalog} 2 showed astrometric offsets for extended, faint objects as large as 0.7 arcsec (Lasker et al. 2008).

In order to further assess the astrometrical calibration of the POSS images acquired via the STScI, large FoV\footnote{field-of-view} images (15 arcmin $\times$ 15 arcmin) around the target were inspected. For the stars on the individual images, the astrometric calibration appears to be satisfactory, within the expected errors. Residual shifts observed in the stellar positions in the subtracted images are observed, but these are caused by proper motions of the stars over the approximate 40-year period. For several galaxies around Leoncino Dwarf (Fig.~3), a simple subtraction procedure has performed well, demonstrating that the astrometric calibration in the extended sources appears also to be satisfactory. Hence, an additional amelioration of the astrometric calibration was not attempted.

Conducive to evidencing any change in the POSS images of Leoncino Dwarf between epochs, the images (POSS I-O and POSS II-J in the blue, and POSS I-E and POSS II-F in the red) were adequately subtracted, a process which requires matching the spatial resolution and the photometry per band in the two epochs. Photometric calibration per band and subtraction of the POSS images was performed according to the following standard prescription (e.g., Annunziatella et al. 2013). Firstly, the POSS images were properly registered, i.e., they were matched in the WCS\footnote{world coordinate system} using Python\footnote{https://www.python.org/} 2.7.5. Aperture photometry of 10 -- 20 stars in each POSS field was performed using ds9\footnote{http://ds9.si.edu/site/Home.html} 7.1, and the average count ratio was used to calibrate the POSS images, assuming that the stars did not vary between epochs and neglecting (small) differences in the filters. The typical normalized rms\footnote{root mean square} deviation of the stellar counts in all four POSS images (red/blue POSS I/POSS II) was 0.3. In the blue and red bands, the normalizations (POSS II to POSS I) are 1.06 and 1.12, respectively. The PyRAF\footnote{http://www.stsci.edu/institute/software\_hardware/pyraf} 2.1.6 package DAOPHOT, specifically the routines DAOFIND (finds the stars in the image), PHOT (performs aperture photometry) and PSF (builds the PSF\footnote{point spread function} model), was then used to extract the PSF in all four images (red/blue POSS I/POSS II). The procedure provided almost circular Gaussian PSFs of FWHM\footnote{full-width half-maximum} 4.7, 3.8, 4.4, and 3.7 arcsec for the blue POSS I, blue POSS II, red POSS I and red POSS II image, respectively. Assuming that the PSF characterizes the POSS images, in order to match the image PSFs, the better resolution image (POSS II) needs to be degraded in resolution by convolving ($\otimes$) it with a Gaussian 2-dimensional kernel (G2DK)

\begin{equation}
{\rm POSS \, II} \otimes {\rm G2DK} = {\rm POSS \, I}.
\end{equation} 

\noindent If the PSFs are Gaussians, then G2DK is also a 2-dimensional Gaussian function with a FWHM that follows from

\begin{equation}
{\rm FWHM}_{\rm POSS \, II}^2 + {\rm FWHM}_{\rm G2DK}^2 = {\rm FWHM}_{\rm POSS \, I}^2.
\end{equation}

\noindent This results in a blue and red G2DK FWHM of 2.7 and 2.4 arcsec, respectively (Eq. [2]). Using Python, the (blue and red) POSS II images were convolved with a G2DK of the corresponding FWHM. The convolved POSS II images (POSS II-J in the blue and POSS II-F in the red), were then subtracted from the POSS I images (POSS I-O in the blue and POSSI-E in the red). Figure~1 contains an example, in the blue, of the PSF variation between epochs, and the effective substraction procedure, on three unsaturated stars in the field, after they have been shifted to account for proper motion between epochs. Figure~2 includes the multi-epoch, multi-band (WCS matched, photometrically calibrated) POSS and POSS convolved images of the target, as well as subtracted and convolved subtracted images. 



\setcounter{figure}{0}

\begin{figure}
\begin{center}

\includegraphics[width=8cm]{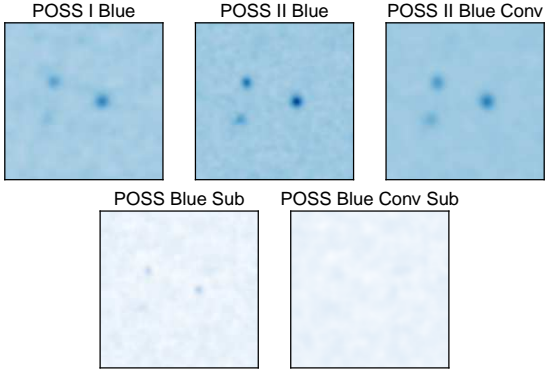}

\caption{{\it Top:} Blue POSS I, blue POSS II, and blue convolved POSS II image of three unsaturated stars in the field, after they have been shifted to account for proper motion. {\it Bottom:} Blue subtracted (POSS II -- POSS I) and blue convolved subtracted (POSS II -- POSS I) image of the same stars. The images illustrate the effectiveness of the subtraction procedure. All images show the same (linear) brightness scale for clarity. All images are 1 arcmin $\times$ 1 arcmin.}

\end{center}
\end{figure}




\setcounter{figure}{1}

\begin{figure}
\begin{center}

\includegraphics[width=9.0cm]{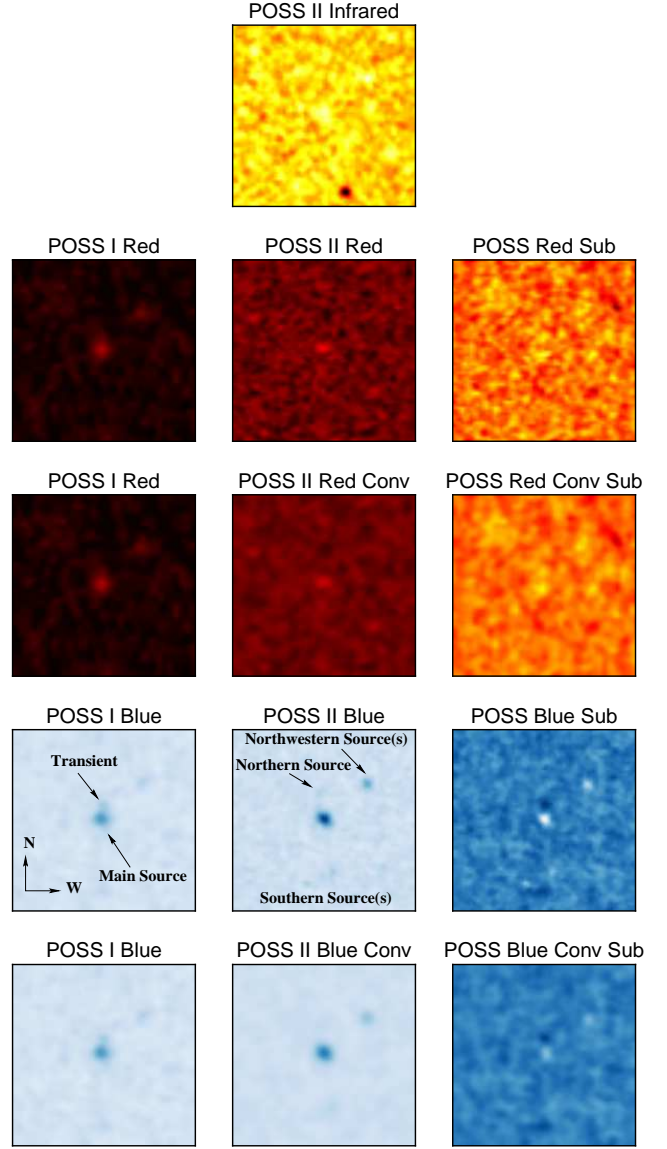}

\caption{{\it Row 1:} Infrared POSS II image. {\it Row 2:} Red POSS I, red POSS II, and red subtracted (POSS II -- POSS I) image. {\it Row 3:} Red POSS I, red convolved POSS II, and red convolved subtracted (POSS II -- POSS I) image. {\it Row 4:} Blue POSS I, blue POSS II, and blue subtracted (POSS II -- POSS I) image. {\it Row 5:} Blue POSS I, blue convolved POSS II, and blue convolved subtracted (POSS II -- POSS I) image. Red/blue (convolved or unconvolved) and subtracted red/blue (convolved or unconvolved) images show the same (linear) brightness scale for clarity. All images are 1 arcmin $\times$ 1 arcmin. Indication of the sources in the FoV is provided in the blue POSS images. Residual signals in the subtracted blue images at the position of the main source are partly an artifact, arising from the imperfect photometric calibration, differences in the PSFs, and partly due to true flux variability. Notice the absence of the transient source in the blue POSS II image.}

\end{center}
\end{figure}


\subsubsection{Transient and Morphological Changes?}

Inspection of the blue POSS I, blue POSS II and blue subtracted images (Fig.~2; row 4) shows that the most accented change is the disappearance, in the second POSS epoch, of the bright blue knot to the North (hereinafter transient source) of the main (host) source; the transient is clearly present in the blue POSS I image (Fig.~2; left-hand column; row 4) at the 5$\sigma$ level (see below), and absent in the blue POSS II higher resolution, higher SNR\footnote{signal-to-noise} image (Fig.~2; middle column; row 4). It is noteworthy that the blue subtracted image (Fig.~2; right-hand column; row 4) mostly removes the main galaxy (see below), leaving a residual signal at the position of the transient, which reinforces the absence of the transient in the blue POSS II image. This variation occurs even when the change in spatial resolution between the two epochs is taken into account (Fig. 2; row 5). The transient source should not be confused with the companion Northern source in Figure~2 (row 4), which is also marked in the HST (Fig.~3; top) and SDSS (Fig.~3; bottom) images. There is no clear evidence for the presence of the transient source in the red POSS I image (Fig.~2; left-hand column; row 2 -- 3), which has an average limiting magnitude of $m_{\rm lim} \approx$ 20.0~mag\footnote{http://www.cadc-ccda.hia-iha.nrc-cnrc.gc.ca/en/dss/?pedisable=true} (see below).

The distance between the peak of the main source and the transient source (blue POSS I) is $\approx$6 pixels or $\approx$6 arcsec ($\approx$0.3 kpc) towards the North, at an angle of $\approx$$-$10$^{\circ}$ (Fig.~2; left-hand column; row 4 -- 5). From the PSF characteristics of the blue POSS I and blue subtracted POSS images (Sect.~2.1.1), the transient source is slightly resolved ($<$ 2 $\times$ PSF FWHM). Because of the faintness of the transient and its proximity to the larger, brighter main source, an automated detection algorithm such as SExtractor (Bertin \& Arnouts 1996), implemented in GAIA\footnote{Graphical Astronomy and Image Analysis Tool} 2016A, was unsuccessful in deblending the two sources. Therefore, a more rudimentary method was employed to determine the magnitude of the transient: using ds9, the transient and the brightest Northwestern source (Sect.~2.2 and 2.3; Fig.~2; left-hand column; row 4 -- 5; Fig~3) were fit with ellipses to perform relative aperture photometry on the photometrically calibrated unconvolved POSS images, after sky subtraction (see below). This procedure shows that the transient source corresponds to $\gtrsim$96\% of the flux of the Northwestern source ($m \approx$ 20.9~mag; USNO B1.0; Monet et al. 2003), which is equivalent to the transient being $\Delta m\lesssim$ 0.05~mag fainter than the Northwestern source (Fig.~2; left-hand column; row 4 -- 5). Similarly, the flux ratio between the main source and the transient is approximately a factor of 6 (Sect.~2.1.4). Given this very simplistic procedure to obtain the transient photometry, the magnitude of the transient is quite uncertain. A magnitude of $m \approx$ 21.0$\pm$0.2~mag will be adopted in the following discussion (Sect.~3), with the confidence interval provided assuming a 20\% error in the flux (detection at a 5$\sigma$ level; see below), corresponding to the estimated photometric error at the maximum intensity extended to the full source, and neglecting unquantifiable systematics errors; the error estimation is included in Appendix A. Favorably, changes in the magnitude as large as $\Delta m \approx$ 1.5 will not significantly alter the conclusions of the present work (Sect.~3). 

The magnitude of the transient puts the transient near the {\em limiting magnitude} of the blue POSS I images commonly quoted in literature ($m_{\rm lim} \approx$ 21.0 -- 21.5~mag; e.g., Djorgovski et al. 2013), which appears associated with the sky brightness level (e.g., Reid et al. 1991). A 9 arcsec $\times$ 13 arcsec elliptical annulus around the main source and transient was used to estimate the local mean value and rms fluctuation of the sky signal. In the case of the transient, the peak flux is found to be $\simeq$5 times the rms fluctuation of the local sky signal. As a sanity check, a similar-sized elliptical annulus around the main source and transient in the blue subtracted image (Fig.~2; right-hand column; row 4) was used to estimate the local rms fluctuation of the sky signal, providing a consistent $\simeq$5 transient peak-to-rms fluctuation value. Therefore, the chance that the transient signal is due to a random fluctuation of the sky signal is as low as 3 $\times$ 10$^{-7}$, assuming the sky noise follows a Gaussian probability density function (similarly small probabilities are found when assuming other probability density functions). Several other independent arguments support that the transient signal is not due to a random fluctuation of the sky signal. For instance, the peak signal of the transient is $\simeq$1.6 higher than the peak of the Northwestern source, which has a (blue POSS I) USNO B1.0 entry (see above), and is clearly detected in the HST (Fig.~3; top) and SDSS (Fig.~3; bottom) images. In addition, the peak sky signals surrounding the galaxy are $\approx$2.5 times smaller than the transient signal.





Because the SNR and resolution of the blue POSS I image is poorer than the blue POSS II image, the transient should be more difficult to detect in the first instance; hence, the fact that the transient is observed in the poorer resolution, poorer SNR image strengthens the reliability of the transient. However, it also raises concerns as to whether it may be an artifact. Various possibilities will be examined and discarded, as follows. A large FoV around the source (15 arcmin x 15 arcmin) was examined, and from the inspection of tens of stars in the field, it was concluded that the transient was not a PSF effect; although the PSF is poorer in the blue POSS I image, it is relatively well-behaved across the field and almost circular (Sect.~2.1.1; Fig. 1). The blue POSS I digitized image does show scratches in several places, but nothing that could explain the transient. Plate distortions would also not produce such a response. Dust on the plate would have created a pinhole effect. There does not appear to be any anomalous optical aberration in the field, nor is there a repeated pattern in other sources. Photographic plate emulsions, such as those used for POSS, typically respond to illumination (i.e., sky) by producing a granular noise structure (Fig.~1 and 2), but the transient signal is clearly distinguishable from the sky signal (see above). Cosmic rays are a possibility, but these photographic plate emulsions are sensitive to less energetic particles that leave a dense signal track. In addition, emulsions were generally thin at the time, so that particles would need to travel almost perfectly parallel to the plane to leave a substantial track. Not only is the transient source only slightly resolved (see above) and unlike a track, inspection of the large FoV around the source does not reveal any such discernible tracks. There is no obvious reason why an anomaly in the telescope, plate, plate emulsion or densiometer would produce such an effect. Hence, the evidence for a true transient is secure.



Residual signals in the blue subtracted images (Fig.~2; right-hand column; row 4 -- 5), at the position of the main source, are partly an artifact, arising from the imperfect photometric calibration (Sect.~2.1.1), differences in the PSFs (Fig.~1), and partly due to true flux variability (Sect.~2.3.1). From the comparison with the PSFs of the POSS images (Sect.~2.1.1), the main source appears resolved. The infrared POSS II image shows only very faint emission at the position of the main source (Fig.~2; row 1). The blue and red POSS I images (Fig.~2; left-hand column; row 2 -- 5) show the main source to be almost circular. Elliptical fits (avoiding the transient source) using ds9 provide axis ratios of $\approx$1. However, the morphology of the main source appears to change in the second epoch POSS images (Fig.~2; middle column; row 2 and 4) to become similar to what is seen in the SDSS image (Fig.~3; bottom); the emission appears to elongate towards the West-Southwest direction. Elliptical fits to the blue and red POSS II images show an axis ratio of $\approx$1.5 and PAs\footnote{position angle} of $\approx$220 and $\approx$1 deg, respectively. Such a change in the morphology of the main source can not be due to effects from the varying PSF (Sect.~2.1.1), nor is it likely due to a systematic astrometric offset (Sect.~2.1.1). It is significant that the morphology change is still apparent in the convolved red/blue POSS II images (Fig.~2; middle column; row 3 and 5). 

A Northwestern source is discernible in the blue POSS I (Fig.~2; left-hand column; row 4 -- 5), blue POSS II (Fig.~2; middle column; row 4 -- 5) and red POSS I images (Fig.~2; left-hand column; row 2 -- 3), corresponding to the brightest of two Northwestern sources apparent in the HST (Fig.~3; top) and SDSS (Fig.~3; bottom) images. A very faint Northern source is apparent in the blue POSS II image (Fig.~2; middle column; row 4 -- 5), as well as in the HST (Fig.~3; top) and SDSS (Fig.~3; bottom) images. A faint Southern source appears in the second epoch blue POSS image (Fig.~2; middle column; row 4 -- 5), discernible also in the HST (Fig.~3; top) and SDSS (Fig.~3; bottom) images.



\subsubsection{Proper Motion and Brightness Centroid Shift?}

Leoncino Dwarf is documented as having a relative proper motion (in RA\footnote{Right Ascension}) of 16 $\pm$ 5 mas yr$^{-1}$ in the USNO B1.0 (Monet et al. 2003) catalog, and an absolute proper motion (in RA) of 7.5 $\pm$ 5 mas yr$^{-1}$ in the PPMXL\footnote{Positions and Proper Motions Extra Large} catalog, relative to a mean epoch of 1979.7 (Roeser, Demleitner \& Schilbach 2010). The errors in the proper motions in Dec\footnote{Declination} in both catalogs are of the order of, or larger, than the proper motions themselves, and so, are not considered further. The PPMXL absolute proper motion value corresponds to $\approx$0.3 arcsec in approximately 40 years, and represents superluminal motion at the distance of Leoncino Dwarf (Table~1). Measuring proper motions beyond 1 Mpc is highly unusual; typically, proper motions are measured only for the Local Group (e.g., van der Marel et al. 2014), with values that are in the range of tens of $\mu$arcsec yr$^{-1}$ (e.g., Brunthaler et al 2007).


As was argued in Section~2.1.1, the PSF is relatively well-behaved, so that it is likely not responsible for a proper motion effect. The astrometrical calibration also appears to be satisfactory, within the expected errors, which should be $\approx$0.3 arcsec for slightly resolved objects such as the transient and main source (Sect.~2.1.1). In order to further investigate this issue, relative astrometry between the main source, companion sources and field sources (seven bright stars and two galaxies within 15 arcmin of the target), was attempted using ds9 and Python. The conclusion is that there does not appear to be any significant relative motion between POSS epochs of the various sources, in the red and blue bands. Hence, it can not be excluded that the apparent flagged proper motion in RA arises from astrometric calibration errors (Sect.~2.1.1). 

It is, however, noteworthy that, in the blue band, the peak of the main source shifts by $\approx$1 pixel ($\approx$1 arcsec) towards the South between POSS epochs (Fig.~2; row 4 -- 5), which is larger than the expected astrometric error in the POSS images (Sect.~2.1.1). This main source brightness centroid shift is not related to PSF variations (Sect.~2.1.1), and no systematic astrometric offset appears to be present (Sect.~2.1.1). It is also not an effect brought on by the disappearance of the transient; the distance from the transient to the main source peak is $\approx$6 arcsec, larger than the typical astrometric error, and the main source-to-transient flux ratio is approximately 6 (Sect.~2.1.1, 2.1.2 and 2.1.4).



\subsubsection{Long-Term 40-year Baseline Variability?}

Rejecting the USNO A2.0 entries, which may have faint-end photometric issues (Sect.~2.1.1), the POSS photometry (Table~2) may suggest that some variation in flux ($\Delta m$ $\lesssim$ 1 mag) has occurred between 1955 and 1998. The photometric accuracy is $\approx$0.3 mag for stellar-like objects, with additional errors for faint, extended objects, which typically may reach $\approx$0.4~mag (Monet et al. 2003; Lasker et al. 2008).

In order to further investigate possible flux variation, the photometrically calibrated unconvolved POSS images were used to investigate variability in the main source using ds9. The comparison between the two epochs appears to demonstrate that the peak of the main source shows a factor $\simeq$2 variation in flux ($\Delta m$ $\simeq$ 0.7 mag), in the blue band, over a period of approximately 40 years, a variation larger than the typical photometric error for faint, extended objects (Lasker et al. 2008). The main source peak appears to have brightened in the blue band, a variation which can also be observed in Figure~2 (middle column; row 4 -- 5). The variation is not an effect of the calibration procedure, as the magnitude of the variation is larger than the calibration applied to the blue POSS I image (Sect.~2.1.1). The brightness increase of the main source peak is also unrelated to the disappearance of the transient; the transient is approximately 6 times lower in flux, and $\approx$ 6 arcsec distant from the main source (Sect.~2.1.2. and 2.1.4). 


\subsection{H16 Data}

H16 contains an ACS\footnote{Advanced Camera for Surveys} HST image of Leoncino Dwarf in the (combined) $\sim$$V$ and $\sim$$I$ filters. The HST images (Fig.~3; top) reveal a blue cluster of bright stars, extended from the Northeast to the Southwest, but more concentrated towards the Southwest, defining the main source. There are also source companions (hereinafter companion sources), all likely galaxies, to the North, Northwest and South of Leoncino Dwarf (Fig.~1; top; Sect.~2.1 and 2.3). The fainter of the Northwestern sources appears to be a red edge-on disk galaxy (Fig.~1; top). The Northern source also appears as an edge-on disk galaxy, but the color is slightly bluer (Fig.~1; top). Both the brighter Northwestern source and the Southern source appear more face-on, and have similar colors to the Northern source (Fig.~1; top). An HI map obtained with the WSRT\footnote{Westerbork Synthesis Radio Telescope}, although of poor resolution ($\sim$22 arcsec  $\times$ $\sim$13 arcsec beam), shows HI emission centered on the optical galaxy, with a hint of an extension towards the South/Southwest and possibly Northwest, towards the companion sources (H16). Although it is not entirely clear from the HI image if there is some association, it appears that, at least the brighter Northwestern and Southern source may be similar redshift companions to Leoncino Dwarf (Sect.~2.1 and 2.3). There are also WIYN\footnote{University of Wisconsin-Madison, Indiana University, Yale University and the National Optical Astronomy Observatories} 0.9~m observations (H16), which show unresolved H$\alpha$ emission to the Southwest. It is this Southwestern HII emission region that was the target for follow-up observations with the KPNO\footnote{Kitt Peak National Observatory} KOSMOS\footnote{Ohio-State Multi-Object Spectrograph} and with the MMT\footnote{Multiple Mirror Telescope} Blue Channel Spectrograph (H16). These spectroscopic observations provide the low-metallicity value for Leoncino Dwarf, as well as other relevant spectroscopic measurements (Table~1; see H16 for details).



\setcounter{figure}{2}

\begin{figure}
\begin{center}

\includegraphics[width=8.0cm]{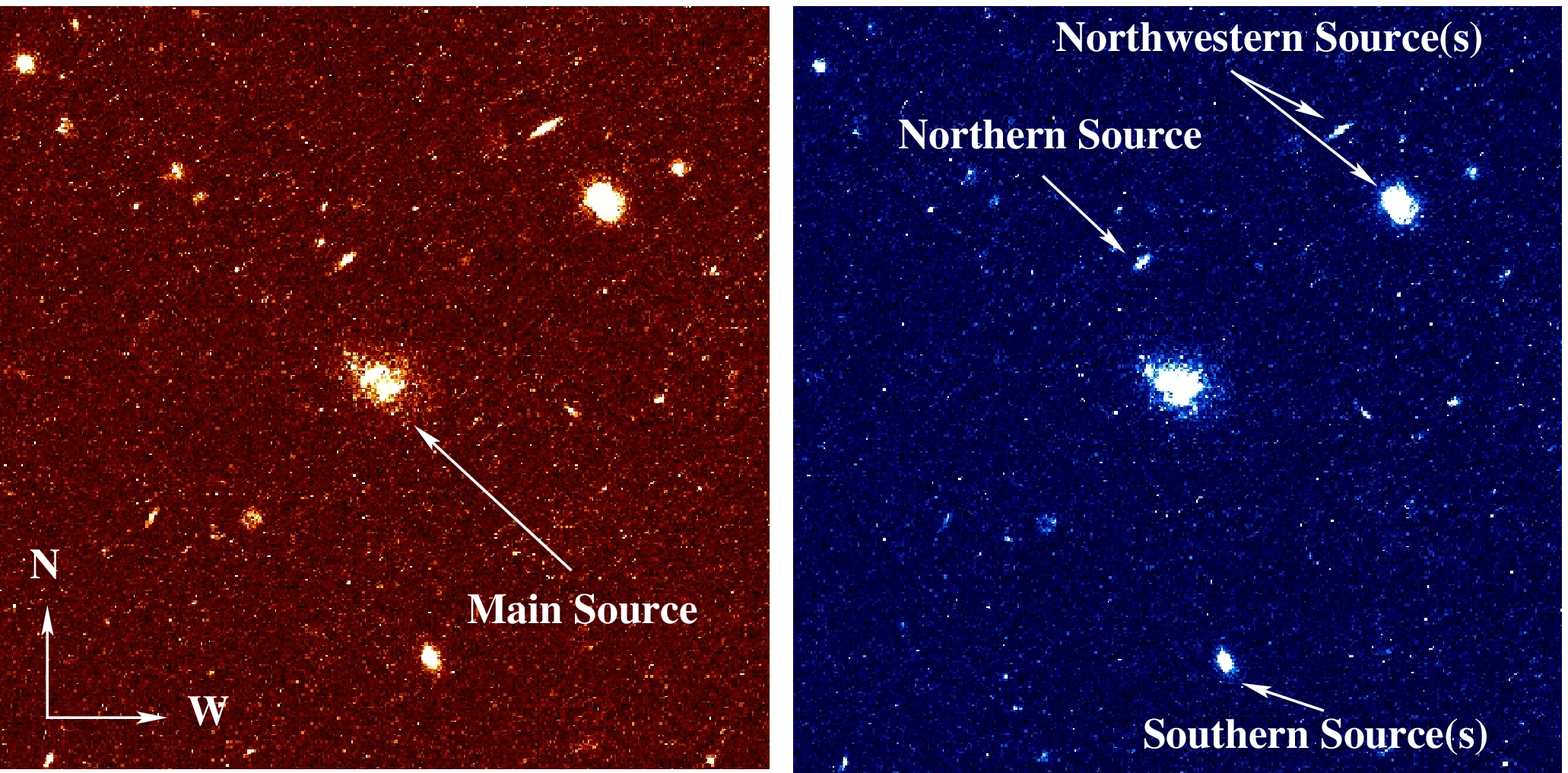}
\includegraphics[width=4cm]{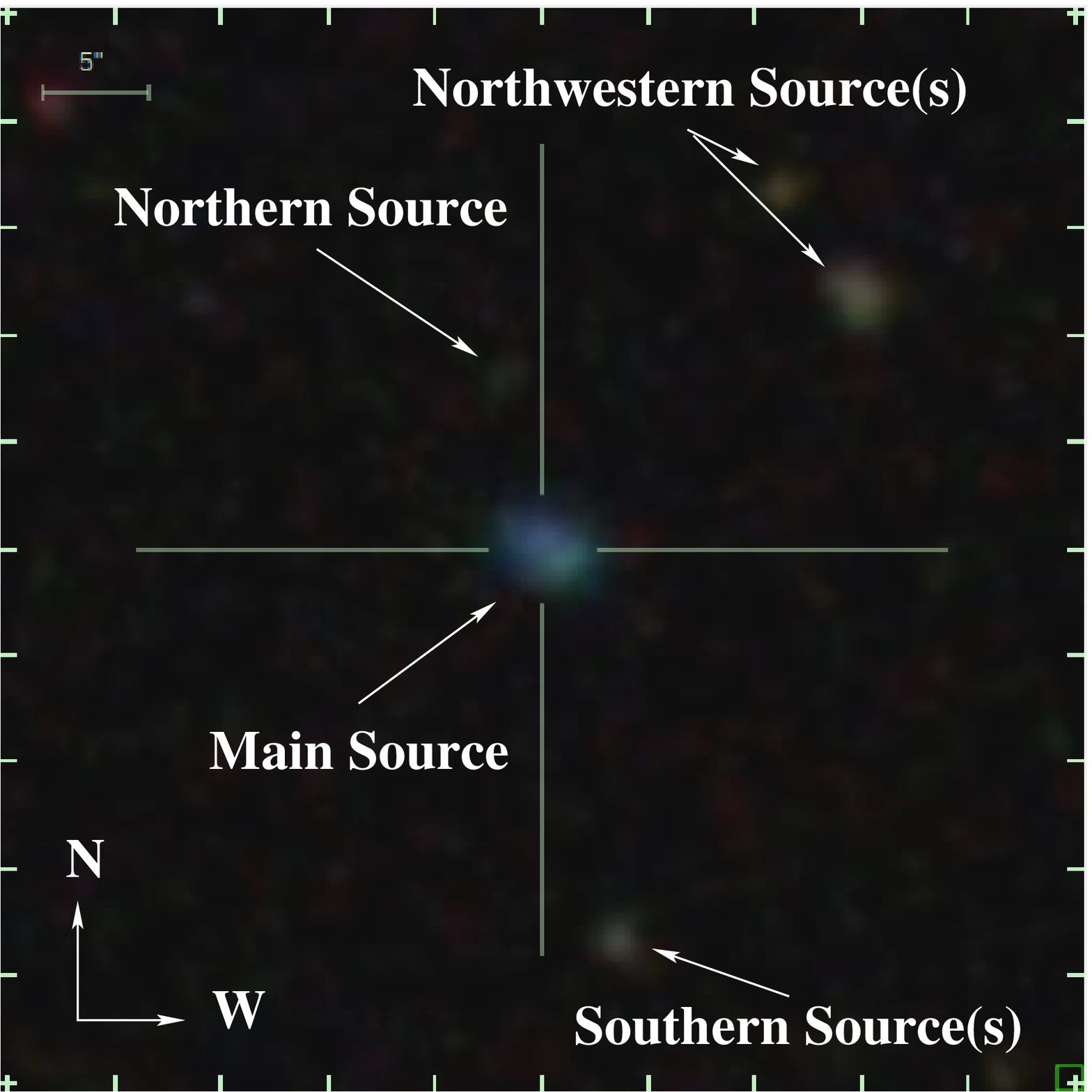}

\caption{{\it Top:} HST $V$-band ({\it left}) and $I$-band ({\it right}) sinh-scale images. {\it Bottom:} SDSS composite image. Notice the companion sources to the North, South and Northwest. The images are 50 arcsec $\times$ 50 arcsec.}

\end{center}
\end{figure}


\subsection{SDSS Data}

Leoncino Dwarf is included in the SDSS photometric catalog, but possesses no SDSS spectroscopy. The SDSS photometric parameters are included in Table~3, and the SDSS composite image is provided in Figure~3 (bottom).



\setcounter{table}{2}

\begin{table}

\footnotesize

\begin{center}

\begin{minipage}{72mm}

\caption{SDSS Photometry of Leoncino Dwarf}

\begin{tabular}{ l | l || l | l }

\hline

Parameter					& Value											& Parameter$^\dagger$			& Value \\

\hline
\hline

$m_{\rm u}$		&	19.79 $\pm$ 0.09 mag		& $R_{\rm u}$	&	4.5 arcsec  \\
$m_{\rm g}$		& 	19.58 $\pm$ 0.04 mag		& $R_{\rm g}$	&	3.6 arcsec \\
$m_{\rm r}$		&	19.55 $\pm$ 0.04 mag		& $R_{\rm r}$ 	&	3.1 arcsec	\\
$m_{\rm i}$		&	19.96 $\pm$ 0.12 mag		& $R_{\rm i}$	&	2.4 arcsec	\\
$m_{\rm	z}$		&	19.86 $\pm$ 0.38 mag		& $R_{\rm z}$	& 	2.9 arcsec	\\

\hline

\end{tabular}

$^\dagger$SDSS Petrosian radius, at 90\% of the light, in the various bands.

\end{minipage}

\end{center}

\end{table}


The SDSS $u$- and $i$-band images essentially probe the main source, which is reproduced also in the $g$- and $r$-band images, albeit as brighter emission (Fig.~3; bottom; Table~3). The SDSS $u$- and $i$-band image of the main source appears to lack some of the extended emission seen to the West-Southwest in the $g$- and $r$-band (Fig.~3; bottom). The main source is barely detectable in the $z$-band. It is to be noted that the $g$- and $r$-band have the least noise; missing weak signals in other SDSS bands are likely due to their higher noise level.

The companion sources observed in the HST image (Fig.~3; top; Sect.~2.2) are also detected in the SDSS image (Fig.~3; bottom). The Northern source possesses no SDSS spectroscopy nor photometry, but shares the same reddish color as the other companion sources. Despite not appearing in the SDSS spectroscopic catalog, the remaining companion sources do appear in the SDSS photometric catalog as galaxies, with magnitudes of $m_{\rm g} \simeq$ 22.5 mag (Southern source), and $m_{\rm g} \simeq$ 21.2 and $m_{\rm g} <$ 22.0 mag (Northwestern sources). 


\subsubsection{Long-Term 60-year Baseline Variability?}
  
Section 2.1.4 discusses possible main source peak variability on an approximately 40-year baseline. Flux variations on an approximately 60-year baseline may be constrained using the color transformation from the SDSS $g$-, $r$- and $i$-bands to the POSS (I-O, I-E, II-J, II-F, II-N) survey filters, transformations which typically have a dispersion of approximately 0.3 mag (Monet et al. 2003). Comparison of the POSS magnitudes (Table~2) and SDSS magnitudes transformed to the POSS filters (Table 4) appear to show some possible variation in the infrared since 1998 ($\Delta m \lesssim$ 1 mag), but a steady blue and red flux since 1998. If the blue POSS main source peak flux variation is real (Sect.~2.1.4), this suggests that the event occurred between 1955 and 1998; however, if the POSS/SDSS infrared flux variation is real, this suggests that the event may still persist (Sect.~2.4.1 and 3).



\setcounter{table}{3}

\begin{table}

\footnotesize

\begin{center}

\begin{minipage}{41mm}

\caption{POSS I/II Magnitudes Corresponding to SDSS Photometry for Leoncino Dwarf}

\begin{tabular}{ l  c  c  c }

\hline

Survey		&	Band$^{\dagger}$	&	$m$ [mag] \\

\hline
\hline
POSS I-O	&	O	 		&	19.7  \\
POSS II-J	&	B$_{\rm J}$	& 	19.6  \\
POSS I-E	& 	E			& 	19.4 \\
POSS II-F	& 	R$_{\rm F}$	&	19.5 \\
POSS II-N	&	I$_{\rm N}$	&	19.6 \\		

\hline

\end{tabular}

$^{\dagger}$See Table~2.

\end{minipage}

\end{center}

\end{table}


\subsection{WISE Data}

Using a 10 arcsec cone search, the WISE\footnote{Wide-Field Infrared Survey Explorer} Post-Cryogenic Single Exposure Source Table reveals three targets (Fig.~4; blue circles), all within $\lesssim$8 arcsec of the galaxy position (Table~5). Four targets (Fig.~4; green circles), all within $\lesssim$8 arcsec of the galaxy position, were also found in the NEOWISE-R\footnote{Near-Earth Object WISE Reactivation} mission Single Exposure Source Table (Table~5). The single exposure data were acquired several months to one year apart (Table~5). One target (Fig.~4; magenta circle) was further found in the AllWISE Reject Table (Table~5). AllWISE is a combination of WISE cryogenic and NEOWISE post-cryogenic phase data, including a recalibration of the original photometry and astrometric solutions integrating the proper motions of reference stars; for high SNR sources in non-confused regions, the astrometric accuracy of AllWISE is $\sim$50 mas radially, relative to the ICRS defined by quasars. The WISE data at 3.35 and 4.6 $\mu$m possess angular resolutions of $\approx$6 arcsec. The SNR of all the targets falls below the value of 5 generally required to be a reliable detection (Table~5). 

\subsubsection{Short-Term Variability?}

Cross-correlation of the WISE Post-Cryogenic/NEOWISE-R/AllWISE data with the blue POSS I image (Fig.~4) shows that one of the NEOWISE-R targets (source 5) is within $\simeq$3 arcsec of the transient source, and another two targets (source 1 and 2) coincide with the main source peak position. There is, however, a difference in magnitude (Table~5) between the NEOWISE-R (source 2) and AllWISE (source 1) target, which exceeds the difference expected from pure noise.

It is unclear whether, and how, the WISE data points and the galaxy are related in terms of position and flux (Fig.~4). There are at least five possibilities for these apparent short-term (several months -- 1 year) changes in magnitude and position: (a) they represent the same source/sources moving across the galaxy (Sect.~2.1.3 and 3), (b) they demonstrate intrinsic variability of the galaxy (Sect.~2.1.4, 2.3.1 and 3), (c) they represent one weak infrared source, likely related to the galaxy itself, but because it is at noise level, the peak will shift slightly between exposures (Sect.~2.1.3), (d) they are spurious signals at noise level, and (e) they correspond to astrometric/photometric shifts due to the new astrometric/photometric calibrations. Given the evidence, it is likely that the WISE detections correspond to astrometric/photometric shifts representing one weak (2 $<$ SNR $<$ 5) infrared source associated with the galaxy. 



\setcounter{figure}{3}

\begin{figure}
\begin{center}

\includegraphics[width=6.0cm]{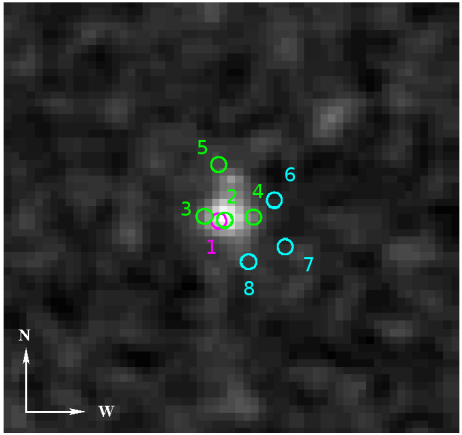}

\caption{Blue POSS I image (linear-greyscale) on which there is superimposed the positions (1 arcsec circles) of the WISE Post-Cryogenic (blue circles), NEOWISE-R (green circles) and AllWISE (magenta circle) targets. All detections have 2 $<$ SNR $<$ 5. One of the NEOWISE-R (source 2) and AllWISE (source 1) targets coincides with the peak in the main source, while one of the NEOWISE-R (source 5) targets is within $\simeq$3 arcsec of the transient source. The image is 1 arcmin $\times$ 1 arcmin.}

\end{center}
\end{figure}







\setcounter{table}{4}

\begin{table}

\scriptsize

\begin{center}

\begin{minipage}{80mm}

\caption{WISE Photometry of Leoncino Dwarf}

\begin{tabular}{ l c c c c c c }

\hline

Survey				&	Source		& Date			&	3.35 $\mu$m		& SNR	& 	4.6 $\mu$m  	&	SNR \\
					& 	[Fig.~4]	&				&	[mag]			&		&	[mag]			& \\

\hline
\hline

ALLWISE	 Reject		& 1				&\ldots			& 18.0		& 3.7		& \ldots		& \ldots	\\	
NEOWISE-R			& 2				& 03/05/2014	& 16.1		& 4.4		& \ldots		& \ldots	\\
					& 3				& 10/11/2014	& 16.8		& 2.2		& \ldots		& \ldots	\\
					& 4				& 10/11/2015	& 15.9		& 3.4		& \ldots		& \ldots	\\
					& 5				& 30/04/2015	& \ldots	& \ldots	& 14.7			& 2.3		\\
WISE Post-Cryo		& 6				& 10/11/2010	& \ldots	& \ldots	& 15.3			& 2.7		\\
					& 7				& 09/11/2010	& 16.6		& 2.8		& \ldots		& \ldots	\\
					& 8				& 10/11/2010	& 16.5	 	& 3.1		& \ldots		& \ldots 	\\

\hline

\end{tabular}

\end{minipage}

\end{center}

\end{table}


\subsection{Other Data}

Other multi-wavelength data for the target was procured via online database inquiries, with searches within a 10 arcsec cone/radius of the target position (Table~1). The source has not been detected in the FIRST\footnote{Faint Images of the Radio Sky at Twenty-Centimeters}, NVSS\footnote{National Radio Astronomy Observatory Very Large Array Sky Survey} nor in the VLSS\footnote{Very Large Array Low-Frequency Sky Survey}. There are no observations of Leoncino Dwarf in the Herschel, Chandra or XMM\footnote{X-ray Multi-Mirror Mission}-Newton database. There also does not appear to be a 2MASS\footnote{Two Micron All-Sky Survey} source associated with the target position. The IAU\footnote{International Astronomical Union} Central Bureau for Astronomical Telegrams\footnote{http://www.cbat.eps.harvard.edu/} was investigated for any SN/nova/unusual variable events at the target position; none were found. The source is not flagged in any of the NEOWISE\footnote{Near-Earth Object Wide-Field Infrared Survey Explorer} mission catalogs as being associated with a Solar System object (Sect.~2.4). The HORIZONS Web-Interface tool\footnote{https://ssd.jpl.nasa.gov/horizons.cgi\#top}, a limited interface to Jet Propulsion Laboratory's HORIZONS system, was used to generate ephemerides for the Solar System planets; no planets were found on the observation date and time at the position of the transient source.


\section{Nature of the Observed Phenomena}

It has been hypothesized that it is probable, from the multi-epoch imaging, that the bright blue knot to the North of the main source is a transient, since it is no longer visible in the blue POSS II (Fig.~2; iddle column; row 4 -- 5) nor in the HST (Fig.~3; top) and SDSS (Fig.~3; bottom) images. During the same approximate 40-year time period, the disappearance of this transient source appears also accompanied by a variation in the main source morphology (Sect.~2.1.2), a change in the brightness centroid of the main source (Sect.~2.1.3), and a peak flux variation of the main source (Sect.~2.1.4 and 2.3.1; Table~2 and 4; Fig.~2; row 4 -- 5). In principle, it could be argued that these phenomenon are uncorrelated. However, several quasi-simultaneously variable, but independent events, in one extragalactic source is highly improbable. Hence, given the timeline and characteristics, it appears more likely that the phenomena are correlated. 

Below, various simultaneous explanations of these observables are attempted, taking into consideration the following criteria, ordered by decreasing reliability in the data: (a) transient source (Sect.~2.1.2), (b) change in the morphology of the main source (Sect.~2.1.2), (c) brightness centroid shift of the main source (Sect.~2.1.3), (d) long-term ($t \lesssim $ 40 year) peak variability of the main source (Sect.~2.1.4 and 2.3.1), and (e) short-term (1 year $\gtrsim t \gtrsim$ several months) variability (Sect.~2.4.1). The scenarios put forward to explain the observations have been ordered according to their ability to account for an increasing number of observables. However, it is to be stressed that this is not synonymous with an increasing probability of the scenario to explain the observations; for example, despite being able to explain most of the observables, the gravitational lens scenario is statistically highly unlikely. Although some of the hypotheses may be clearly discarded, several others are still viable under certain specific conditions.  Table~6 provides a compilation of the various scenarios that have been explored, and the observables they are able to reproduce.


Throughout, a distance of $D \simeq$ 8~Mpc to Leoncino Dwarf (Table~1), and an apparent magnitude for the transient of $m \approx$ 21.0 mag (Sect.~2.1.2), have been adopted. However, the distance (H16) and magnitude (Sect.~2.1.2) are uncertain. If the apparent magnitude of the transient is fainter by $\Delta m \approx$ 1.5~mag, either due to doubling the distance to Leoncino Dwarf ($D \approx$ 16~Mpc) or due to a large error in the estimated transient photometry, the following results are not substantially altered.



\setcounter{table}{5}

\begin{table*}

\scriptsize

\begin{center}

\begin{minipage}{150mm}

\caption{Nature of the Observed Phenomena}

\begin{tabular}{ l c c c c c c c c c c c c }

\hline

Observable			& Stellar 	  &  Dust			&	Stellar & Solar System	& Accretion 	& Hypernova/ & Stellar  & Nova	& LBV & Gravitational \\
					& Variability & Enshroudment	&   Cluster	& Object		& Variability	& SN		 & Merger   &        &  Star & Lensing       \\

\hline
\hline

Transient Source			&\xmark &  \xmark		& \cmark?	& \cmark?	& \cmark?	& \cmark	& \cmark	 	& \cmark	& \cmark & \cmark 	\\
Morphological Changes		&\cmark &  \xmark		& \xmark	& \xmark	& \xmark	& \xmark	& \cmark?		& \cmark?	& \cmark? & \cmark	\\
Brightness Centroid Shift 	&\cmark &  \xmark		& \xmark	& \xmark	& \xmark	& \xmark	& \cmark?		& \cmark?	& \cmark? & \cmark  \\
Long-term Variability		&\cmark &  \xmark		& \xmark	& \xmark	& \xmark	& \xmark	& \xmark	 	& \xmark	& \xmark & \cmark	\\
Short-term Variability		&\cmark &  \xmark		& \xmark	& \xmark	& \xmark	& \xmark	& \xmark		& \xmark	& \xmark & \cmark?	\\

\hline

\end{tabular} 
'?' Refers to an uncertain, improbable or highly contrived explanation.

\end{minipage}

\end{center}

\end{table*}


\vspace{0.2cm}

\noindent{\bf \em Stellar Variability in the Main Source:} As a low-surface-brightness galaxy (Table~1), the shift in the brightness centroid (Sect.~2.1.3), the morphology change (Sect.~2.1.2), and the (long- and short-term) flux variation in the main source (Sect.~2.1.4, 2.3.1 and 2.4.1), may be due to the variability ($\Delta m \simeq$ 0.7~mag, the main source peak variability in the blue) of a few individual massive stars in the galaxy, stars that are clearly resolved in the HST images (Fig.~3; top). However, such a scenario requires an independent explanation for the transient source (Sect.~2.1.2).

\vspace{0.2cm}

\noindent {\bf \em Dust Enshroudment by Feedback Processes:} In galaxies there are two types of feedback processes to consider: those related to massive black holes (M$_{\rm BH} \gtrsim$ 10$^6$ M$_{\odot}$), and those related to star formation. There is evidence for relatively strong SN-driven winds ($v \simeq $ 100 -- 400 km s$^{-1}$) originating from HII regions (e.g., Olmo-Garc\'\i a et al. 2017 and references therein). Massive black holes, on the other hand, show a large range in outflow velocities, ranging from $v \simeq $ 200 km s$^{-1}$ in quiescent galaxies (e.g., Cheung et al. 2016), to velocities of thousands of kilometers per second in quasars (e.g., Rogerson et al. 2015). In principle, winds/outflows can sweep up dusty gas around bright sources, thus producing optical transients.

In approximately 40 years, a black hole-driven outflow ($v \simeq$ 10~000 km s$^{-1}$) can travel a radial distance of $r \approx$ 0.4 pc, while SN-driven winds ($v \simeq$ 100 km s$^{-1}$) from an HII region can travel a radial distance of $r \approx$ 0.004 pc. However, such distances are very small when compared with the overall size of Leoncino Dwarf ($s \approx$ 0.4 kpc, for $D \simeq$ 8 Mpc; Table~1), and with the distance between the transient source and peak of the main source ($d \approx$ 0.3 kpc, for $D \simeq$ 8 Mpc; Table~1; Fig.~2 and 3). Hence, dust-enshroudment by winds/outflows from massive black holes and/or by SN is likely not the cause of the disappearance of the transient source (Sect.~2.1.2), nor the explanation for the change in morphology/brightness centroid shift/variability (Sect.~2.1.2, 2.1.3, 2.1.4 and 2.3.1).

\vspace{0.2cm} 

\noindent {\bf \em Fading Stellar Cluster:} The transient source may be a stellar cluster that fades with time. If the transient source is a young stellar cluster, commonly associated with star-forming regions, the cluster ($m \approx$ 21.0~mag; Sect.~2.1.2) may contain anywhere from 10 to 25 O-type stars (e.g., Walborn et al. 2002). However, this scenario can be abondoned (e.g., Allen et al 2007); it is inconsistent with the isolation, with the lack of a discernible HII region at the position of the transient (Fig.~2 and 3), with the offset of the transient relative to the main source ($d \approx$ 0.3 kpc, for $D \simeq$ 8 Mpc; Table~1), and with the (quasi-simultaneous) disappearance of a large number of stars in approximately 40 years (Fig.~2; row 4 -- 5), either due to their lifespan (a typical O-type star lifetime is 3 -- 6 Myr) or due to dust enshroudment (see also {\em Dust Enshroudment by Feedback Processes}). In addition, it does not provide an explanation for the other observables (Sect.~2.1.2, 2.1.3, 2.1.4 and 2.3.1).

The absolute magnitude of a globular cluster is $M \simeq -$10 -- $-$5~mag, which, for a distance of $D \simeq$ 8 Mpc (Table~1), translates into an apparent magnitude of $m \simeq$ 19.5 -- 24.5~mag  (e.g., Brodie \& Strader 2006); hence, the optical magnitude of the transient ($m \approx$ 21.0~mag; Sect.~2.1.2) is consistent with a globular cluster. As globular clusters can be found in halos (e.g., Brodie \& Strader 2006), this scenario could explain the isolation, the lack of a host HII region (Fig.~2 and 3) and the offset of the transient relative to the main source ($d \approx$ 0.3 kpc, for $D \simeq$ 8 Mpc; Table~1). However, the size of the transient region appears inconsistent with typical globular cluster sizes (10 -- 30 pc; Brodie \& Strader 2006). In addition, such a scenario can be discarded based on the impossibility of the globular cluster disappearance over a period of approximately 40 years (Fig.~2; row 4 -- 5; see also {\em Dust Enshroudment by Feedback Processes}), as well as not providing an explanation for the other time-dependent observables (Sect.~2.1.2, 2.1.3, 2.1.4 and 2.3.1).

\vspace{0.2cm}

\noindent {\bf \em Passing of a Solar System Object:} The transient source may be associated with the passing of a Solar System object near Leoncino Dwarf, providing an explanation for the isolation, the lack of a host HII region (Fig.~2 and 3), the transient/main source offset ($d \approx$ 0.3 kpc, for $D \simeq$ 8 Mpc; Table~1) and the lack of a radio signature (Sect.~2.5). An asteroid at a distance of $D \approx$ 1 AU, with an apparent magnitude of $m \approx$ 21.0 mag (Sect.~2.1.2), and a typical albedo value (0.05 -- 0.25), would possess a size anywhere from $\approx$150 to $\approx$350~m\footnote{https://cneos.jpl.nasa.gov/tools/ast\_size\_est.html}. Generally, these asteroids possess orbital periods between 3 and 6 years, and proper motions in the range of several arcsec per minute. However, this scenario has several severe issues. Firstly, it would imply that the blue POSS I image (Fig.~2; left-hand column; row 4 -- 5) was acquired during the small time window (several minutes) of the object passage near the galaxy. Secondly, Solar System objects are generally brighter in the red band than in the blue band (e.g., Juri\'c et al. 2002), which contradicts the observational data for the galaxy (Fig.~2; left-hand column; row 2 -- 5; Table~2). Thirdly, Solar System objects are periodical/quasi-periodical, so that they are fairly well-documented; the fact that there is no registered object implies that it would have to be a new, undocumented Solar System object (Sect.~2.5). Lastly, such a scenario can not simultaneously explain the transient source/varying morphology/brightness centroid shift/flux variability (Sect.~2.1.2, 2.1.3, 2.1.4 and 2.3.1) unless multiple objects/events are involved. Given the above cumulative evidence, it is highly unlikely that the event is Solar System object-related.

\vspace{0.2cm} 

\noindent{\bf \em Variable Accretion Onto a Compact Object:} Because dwarf galaxies are chemically and dynamically unevolved, they possess the ideal conditions for the presence of IMBHs\footnote{intermediate mass black holes} of masses M$_{\rm BH} \simeq 10^4$ -- $10^6$ M$_{\odot}$ (e.g., Izotov, Thuan \& Guseva 2007; Izotov \& Thuan 2008; Reines et al. 2013), formed during the early phases of the Universe. Black hole-related variability, occuring at optical wavelengths and on different timescales, is associated with the accretion disk and may have several origins, ranging from disk instabilities (e.g., Siemiginowska \& Elvis 1997), to variations in the accretion rate/mode (e.g., Zuo, Lui \& Jiao 2012), the tidal disruption of nearby stars (e.g., Bogdanovi\'c et al. 2004), and X-ray re-processing (e.g., McHardy et al. 2016). X-ray binaries are also accretion-powered sources, where a donor (star) transfers material to an accretor, generally a neutron star or a (stellar-mass) black hole (e.g., Casares, Jonker \& Israelian 2017). X-ray binaries typically show optical counterparts (the accretion disk or the star itself), with absolute magnitudes ranging from $M \approx$ $-$10 -- 5 mag (LMXBs\footnote{low-mass X-ray binaries} and microquasars) to $M \approx -$5 -- 5 mag (HMXBs\footnote{high-mass X-ray binaries}), depending on the mass of the donor (e.g., Casares, Jonker \& Israelian 2017). They are also variable on different (shorter) timescales, showing erratic light curves, pulsations, quasi-periodic oscillations and transient accretion events (e.g., Casares, Jonker \& Israelian 2017). 

If the observed transient source is associated with an accreting object in Leoncino Dwarf, an upper limit to its mass (M$_{\rm acc}$) and accretion rate ($\dot{\rm M}_{\rm acc}$) can be estimated for a thin-disk approximation and an 0.1 efficiency, assuming the accretion is Eddington-limited. The observed magnitude of the transient source ($m \approx$ 21.0~mag; Sect.~2.1.2) then provides an accretor mass of M$_{\rm acc} <$ 50 M$_{\odot}$ and an accretion rate of $\dot{\rm M}_{\rm acc} <$ 10$^{-7}$ M$_{\odot}$ yr$^{-1}$; such values are within the realm of LMXBs (e.g., Casares, Jonker \& Israelian 2017). Indeed, a subgroup of LMXBs are transient sources, as a result of accretion instabilities; they are generally undetectable in the optical (and X-rays), but become observable for a period of a few days (e.g., Casares, Jonker \& Israelian 2017). As LMXBs can be commonly found in globular clusters (e.g., Casares, Jonker \& Israelian 2017), which can, in turn, occur in halos (see also {\em Fading Stellar Cluster}), such a scenario is not inconsistent with several of the characteristics (magnitude, isolation, lack of a host HII region and transient/main source offset; Fig.~2 and 3; Sect.~2) of the transient source. However, this would imply that the blue POSS I image (Fig.~2; left-hand column; row 4 -- 5) was acquired during that brief, bright period. In addition, an accreting object at the position of the transient does not provide an explanation for the other observables (Sect. 2.1.2, 2.1.3, 2.1.4 and 2.3.1). Therefore, this scenario is deemed highly unlikely.

\vspace{0.2cm}

\noindent{\bf \em Hypernova/SN:} Hypernovae are generally associated with stellar explosions, corresponding to some of the most catastrophic events in the Universe. Models for hypernova are diverse and include black holes, magnetars, pair-instabilities in massive, low-metallicity stars and stellar systems in unusual configurations (e.g., Nomoto et al. 2004). Hypernovae are also known to be associated with long-duration GRBs, which appear to be predominantly hosted by metal-poor, actively star-forming, subluminous galaxies (e.g., Modjaz et al. 2008; Lyman et al. 2017). Hypernovae generally peak at an absolute magnitude of $M \simeq -$21 -- $-$20~mag, which corresponds to an apparent magnitude of $m \simeq$ 8.5 -- 9.5~mag, for a distance of $D \simeq$ 8 Mpc (Table~1; e.g., Maeda et al. 2003). There is a quick drop-off of $\Delta m \approx$ 1~mag in the first 30 -- 50 days (e.g., Maeda et al. 2003). 

Type Ia SN occur in binary systems in which one star is a white dwarf, as a result of the explosion of the white dwarf via the accretion mechanism; these SN constitute a hetergeneous class of sources (e.g., Maoz, Mannucci \& Nelemans 2014). Type Ia SN are typically hosted by normal field galaxies, showing no preference for an association with star-forming regions (e.g., Childress et al. 2013). The light curve for a typical type Ia SN shows a peak at an absolute magnitude of $M \simeq -$20 -- $-$ 19~mag, which corresponds to an apparent magnitude of $m \simeq$ 9.5 -- 10.5~mag, for a distance of $D \simeq$ 8 Mpc (Table~1; e.g., Firth et al. 2015). There is a quick magnitude drop-off of $\Delta m \approx$ 1~mag in the first 30 days (e.g., Firth et al. 2015). Peculiar, lower luminosity, quicker drop-off, type Ia SN have been documented, with typical peak absolute magnitudes of $M \simeq -$17~mag (SN2002cx prototype; Jha et al. 2006), the most extreme of which was found to have a peak absolute magnitude of and $M \simeq -$14~mag (e.g., Foley et al. 2009), corresponding to apparent magnitudes of $m \simeq$ 12.5 and 15.5~mag, respectively. 

Type II SN, expected in a system such as Leoncino Dwarf with signs of ongoing star formation, are the result of the rapid collapse, and subsequent violent explosion, of a massive (M$_{\star} \simeq $ 8 -- 50 M$_{\odot}$) star (e.g., Smartt et al. 2009). Typically, type II SN show a peak absolute magnitude of $M \simeq -$17~mag, which, for a distance of $D \simeq$ 8 Mpc (Table~1), translates into an apparent magnitude of $m \simeq$ 12.5~mag (e.g., Kasen \& Woosley 2009). For approximately 100 days after the initial explosion, the magnitude of a type II SN is maintained within $\Delta m \approx$ 1~mag of the peak magnitude, after which the magnitude drops; 5 months after the explosion the absolute magnitude is still $M \approx -$15 mag (e.g., Kasen \& Woosley 2009). 

Calcium-rich SN are rare, peculiar, low-luminosity SN which show strong calcium lines, approximately two months after their peak brightness (SN2005E prototype; e.g., Perets et al. 2010). They are more prevalent in early-type hosts in dense environments and/or showing signs of a recent merger, and are often found offset from the galaxy center, sometimes as far as 150 kpc (e.g., Foley 2015). The origin of such SN remains controversial. The lack of a post-explosion source rules out massive star progenitors and globular clusters, as well as compact dwarf galaxy hosts, unless they are ultra-faint dwarf satellites (e.g., Lyman et al. 2016). The large offsets suggest kicked, high-velocity systems, such as old, merging white dwarf -- neutron star binaries or the helium detonation in double white dwarf systems interacting with a supermassive black hole (e.g., Foley 2015; Lyman et al. 2016; Lunnan et al. 2017). Calcium-rich SN are generally less luminous then other SN and are faster fading, dropping off by $\Delta m \approx$ 1 mag within the first month (e.g., Perets et al. 2010). A typical peak absolute magnitude of $M \simeq -$15~mag corresponds to an apparent magnitude of $m \simeq$ 14.5~mag (for a distance of $D \approx$ 8 Mpc; Table~1; e.g., Foley et al. 2015).

The transient source is much fainter ($m \approx $ 21.0~mag; Sect.~2.1.2) than would be expected from a hypernova/SN observed at peak, at this distance (Fig.~2; left-hand column; row 4 -- 5). If the transient is a hypernova/SN, its apparent magnitude ($m \approx $ 21.0~mag; Sect.~2.1.2) suggests that it was detected quite some time after its initial explosion. A transient arising from a core-collapse SN, associated with regions of massive star formation (e.g., Smartt et al. 2009), is inconsistent with the transient/main source offset ($d \approx$ 0.3 kpc, for $D \simeq$ 8 Mpc; Table~1), the isolation and the lack of a host HII region (Fig.~2 and 3). On the other hand, a calcium-rich SN origin for the transient (e.g., Perets et al 2010; Foley et al. 2015; Lyman et al. 2016; Lunnan et al. 2017) is compatible with the (possible) interaction of the host with companion galaxies (Sect.~2.2), the isolation, the lack of a post-explosion source, the lack of a host HII region (Fig.~2 and 3) and the transient/main source offset ($d \approx$ 0.3 kpc, for $D \simeq$ 8 Mpc; Table~1). Notwithstanding, because the hypernova/SN was observed during a very late stage, a (net) magnitude variation of the transient of $\Delta m \approx$ 1 -- 2 mag over approximately 60 years should have produced a detectable source in the HST images (Fig. 3; top; ACS limiting magnitude\footnote{http://www.stsci.edu/hst/acs/documents/handbooks/current/c05\_imaging3.html} for 1000 s integration is $m_{\rm lim} \approx$ 26 -- 27.5 mag). In addition, the role of the hypernova/SN in the changing morphology/brightness centroid shift/flux variability (Sect.~2.1.2, 2.1.3, 2.1.4 and 2.3.1) remains unclear. Hence, although this scenario can not be entirely discarded, it does pose challenges to explain several of the observables.

\vspace{0.2cm}

\noindent{\bf \em Stellar Merger:} Stellar mergers generally produce events with complex, irregular, multi-peak light curves (e.g., Smith et al. 2016). Within 1 -- 2 months, light curves may show magnitude variations as large as $\Delta m \approx$ 5 mag (e.g., Smith et al. 2016). Peak absolute magnitudes range from $M \simeq -$10 -- $-$5, depending on the mass of the progenitor stars (e.g., Kochanek, Adams \& Belczynski 2014), which translates into apparent magnitudes of $m \simeq$ 19.5 -- 24.5~mag, for a distance of $D \simeq$ 8 Mpc (Table~1). It is, therefore, possible that the transient source ($m \approx$ 21.0~mag; Sect.~2.1.2) was a stellar merger, observed at, or near, peak magnitude (Fig.~2; left-hand column; row 4 -- 5). In addition, (net) fading of the stellar merger peak event by $\Delta m \approx$ 5 mag (over a period of approximately 40 years) may possibly produce an apparent brightness centroid shift/morphology variation in the main source (Sect.~2.1.2 and 2.1.3). Because stellar merger events can be quite heterogenous (e.g., Smith et al. 2016), this scenario is not inconsistent with the isolation, the lack of a host HII region (Fig.~2 and 3) and the transient/main source offset ($d \approx$ 0.3 kpc, for $D \simeq$ 8 Mpc; Table~1). However, this scenario can not readily account for the peak flux variability in the main source (Sect.~2.1.4 and 2.3.1). Hence, as a stellar merger potentially explains most of the observables, it remains as a possible scenario for the observed phenomenon.

\vspace{0.2cm}

\noindent {\bf \em Nova:} Novae, possible progenitors of type Ia SN, consist of the cataclysmic thermonuclear explosion of an accreting white dwarf in a binary system, and may be recurrent. Absolute magnitudes of a nova are $M \simeq-$10 -- $-$5~mag at peak, which, for a distance of $D \simeq$ 8 Mpc (Table~1), translates into an apparent magnitude of $m \simeq$ 19.5 -- 24.5~mag (e.g., Tang et al. 2014). Nova generally show a light curve with a flatter decline after peak magnitude than a type II SN; a fast nova will typically take less than 25 days to decay from peak by $\Delta m \approx$ 2 mag, while a slow nova will take over 80 days (e.g., Hachisu \& Kato 2015). Because the magnitude of the transient source is $m \approx$ 21.0 mag (Sect.~2.1.2), this signals that the transient could have been a nova observed at, or near, peak magnitude (Fig.~2; left-hand column; row 4 -- 5). As some novae can be found in galaxy outskirts (e.g., Shafter et al 2014), this scenario is not inconsistent with the isolation, the lack of a host HII region (Fig.~2 and 3) and the transient/main source offset ($d \approx$ 0.3 kpc, for $D \simeq$ 8 Mpc; Table~1). (Net) fading of the nova by $\Delta m \approx$ 4 mag after peak over a period of approximately 40 years can possibly provide an explanation for the brightness centroid shift/morphology variation in the main source (Sect.~2.1.2 and 2.1.3), but, like the stellar merger scenario, it can not easily account for the peak flux variability in the main source (Sect.~2.1.4 and 2.3.1). Therefore, this scenario remains as a possible explanation, as it describes most of the observables.

\vspace{0.2cm}

\noindent{\bf \em LBV\footnote{luminous blue variable} Star:} LBVs are evolved supergiant/hypergiant stars, some of the most variable, luminous and massive (20 -- 100 M$_{\odot}$) stars observed. These sources belong to the S Doradus instability strip of the Hertzsprung-Russell diagram. The high mass loss rates and high luminosities result in short lifetimes (e.g., Groh et al. 2014) for the progenitor (few Myr) and LBV phases ($<$ Myr). The traditional view of LBV stars is that they correspond to a transitional phase of the most massive single stars (e.g., Humphreys \& Davidson 1994). More recently, and largely based on population comparison and on the isolated environments of some LBV stars, it has been suggested that LBV stars are the product of binary evolution, i.e., they are evolved massive blue stragglers (e.g., Smith \& Tombleson 2015; Smith 2016). LBV stars are rare, and only several tens of these sources are known in the Milky Way (e.g., Naz\'e, Rauw, \& Hutsem\'ekers 2012) and Local Group galaxies (e.g., Massey 2010). 

LBV stars show unpredictable, and sometimes dramatic, variations in both their brightness and their spectra. LBV stars can be quiescent or dormant for decades or centuries, during which they are generally of spectral type B. Variations of $\Delta m \approx$ 0.1 -- 0.2 mag can occur on timescales of a day to several weeks/months (e.g., Sterken 2003). Intermittent variations of $\Delta m \lesssim$ 2 mag, where the bolometric luminosity remains constant but the spectral type varies from early B supergiant to late B/early A, can occur on timescales of years to a decade (e.g., Sterken 2003). Very rare 'giant eruptions', sometimes mimicking SN events ('SN imposters'; e.g., Kochanek, Szczygie\l~\& Stanek 2012), can occur on timescales of several tens of years, and can produce magnitude variations of $\Delta m \approx$ 3 mag, accompanied by large mass loss rates (with the possible production of nebulae), and an increase in the bolometric luminosity (e.g., Sterken 2003). 

Among XMPs, there is at least one documented case of a giant eruption associated with a LBV star: DDO~68 (Pustilnik et al. 2008; Izotov \& Thuan 2009; Bomans \& Weis 2011; Pustilnik et al. 2017). The eruption appears to have lasted for 2 -- 6 years; between 2005 and 2010 the LBV star brightened by $\Delta m \approx$ 3.1 mag, while between 2010 and 2015 the LBV star dimmed by $\Delta m \approx$ 3.7 mag, with a maximum absolute magnitude of $M \approx$ $-$10.5 mag (Pustilnik et al. 2017). Archival (POSS) data for DDO~68 appear to suggest possible previous eruptions in 1955 and 1999 (Bomans \& Weis 2011; Pustilnik et al. 2017). 

Utilizing the full range of peak absolute magnitudes observed in LBV stars ($M \simeq$ $-$13 -- $-$9 mag; e.g., Smith et al. 2011b) provides a range of peak apparent magnitudes of $m \simeq$ 16.5 -- 20.5 mag, for a distance of $D \simeq$ 8 Mpc (Table~1). Hence, if the transient source ($m \approx $ 21.0~mag; Sect.~2.1.2) is a LBV star, it must have been observed a short time after its peak (Fig.~2; left-hand column; row 4 -- 5); quiescent LBV stars can have absolute magnitudes as low as $M \simeq$ $-$6 mag (e.g., Smith et al. 2011b), or $m \simeq$ 23.5 mag (for a distance of $D \simeq$ 8 Mpc; Table~1). A LBV star provides an adequate explanation for the transient timescale, the isolation (e.g., Smith \& Tombleson 2015; Smith 2016), the lack of a host HII region (Fig.~2 and 3; see below, however) and the transient/main source offset ($d \approx$ 0.3 kpc, for $D \simeq$ 8 Mpc; Table~1; see below, however). In addition, (net) fading of the LBV peak event by $\Delta m \lesssim$ 3 mag (over a period of approximately 40 years) may possibly produce an apparent brightness centroid shift/morphology variation in the main source (Sect.~2.1.2 and 2.1.3). However, this scenario has its caveats. Firstly, it is difficult to interpret the flux variability of the main source (Sect.~2.1.4 and 2.3.1) within such a context. Secondly, although a fraction of LBV stars are isolated (e.g., Smith \& Tombleson 2015; Smith 2016), in the only XMP with a documented LBV star, the LBV star appears embedded within an HII region (DDO~68; Pustilnik et al. 2017); no clear HII region at the location of the transient source is discernible in the HST images (Fig.~3; top), although it should be detectable given the typical lifetime of an HII region (few Myr; e.g., Alvarez et al. 2006). Thirdly, a (net) magnitude variation of $\Delta m \lesssim$3 mag of the transient over a period of approximately 60 years would result in a quiescent LBV source ($m \lesssim$ 23.5 mag) that should have been detectable in the HST images (Fig. 3; top). Lastly, as LBV stars brighten, they become redder (e.g., Sterken 2003), which appears to contradict the POSS data (Table~2; Fig.~2; row 2 -- 5). Consequently, as the LBV scenario explains some of the observables challenged by other scenarios (e.g., transient timeline), it remains a contender for the observed phenomenon.
 
\vspace{0.2cm}

\noindent {\bf \em Gravitational Lensing:} The transient source and/or change in morphology/brightness centroid shift/flux variation may be due to a lensing effect (Sect.~2.1.2, 2.1.3, 2.1.4 and 2.3.1). Figure 5 contains plots of the gravitational lensing equations in Appendix B for a range of distances and lens masses. 



\setcounter{figure}{4}

\begin{figure*}	
\begin{center}

\includegraphics[width=6.5cm]{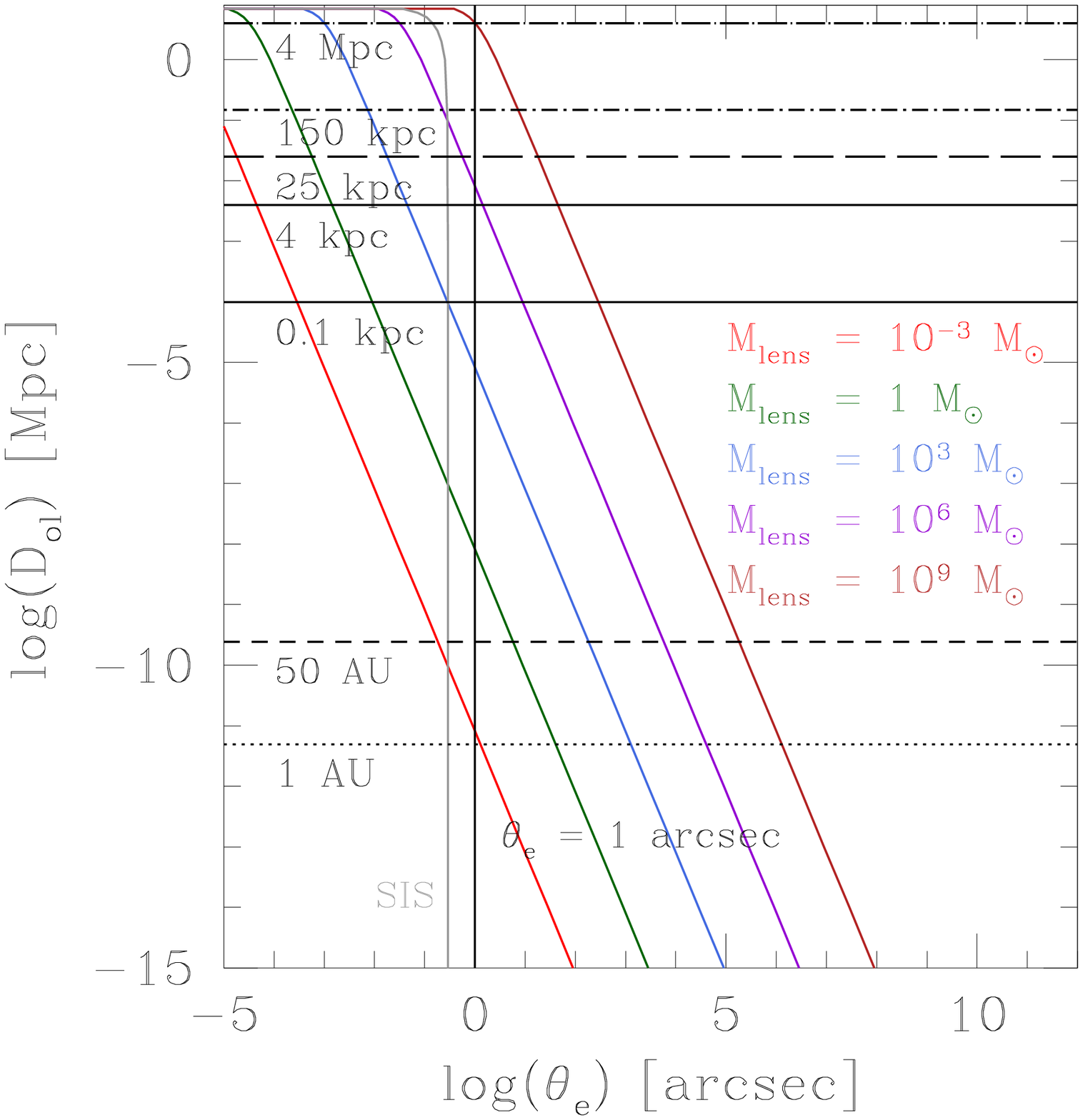}
\includegraphics[width=6.5cm]{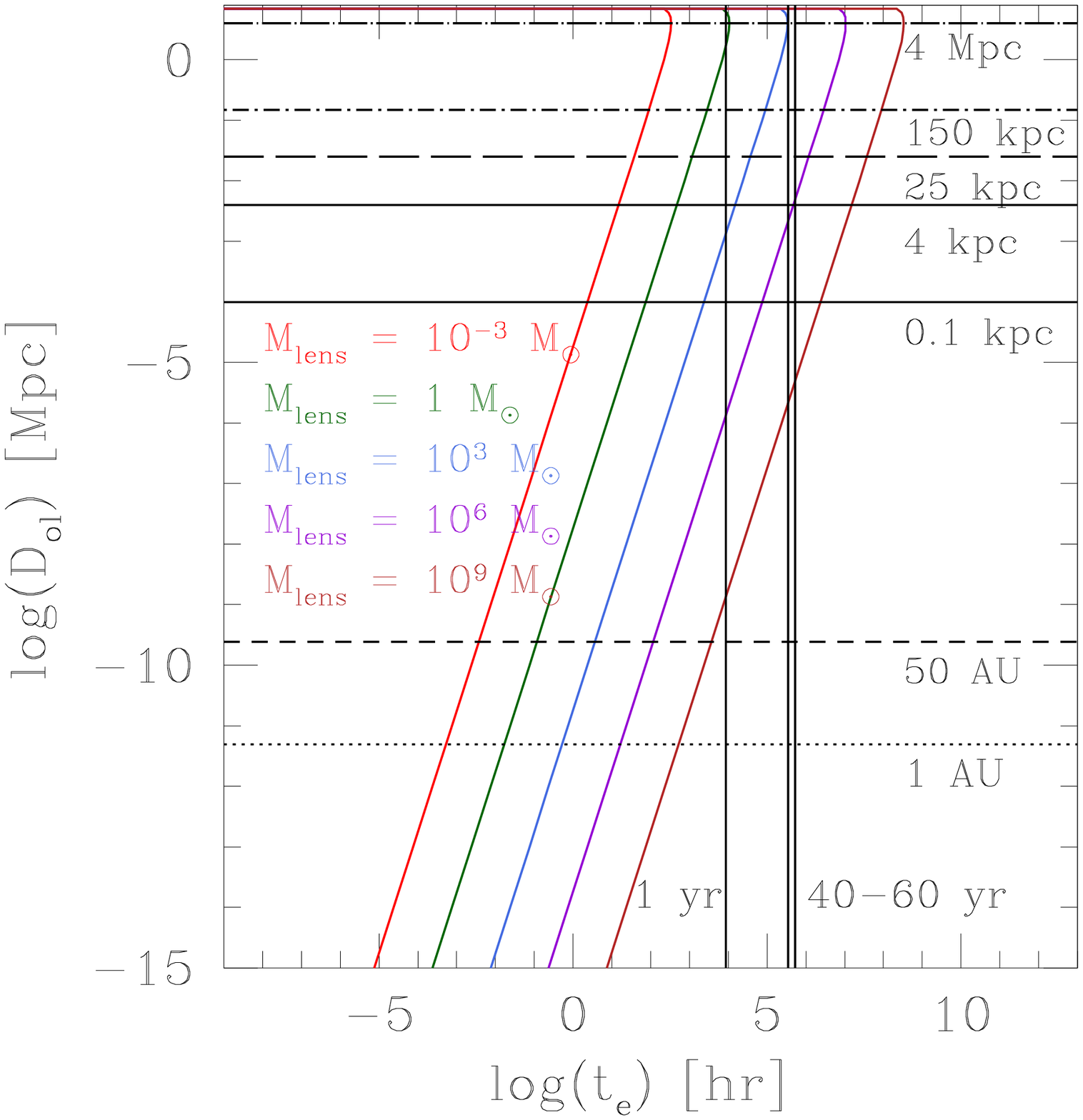}
\\
\includegraphics[width=6.5cm]{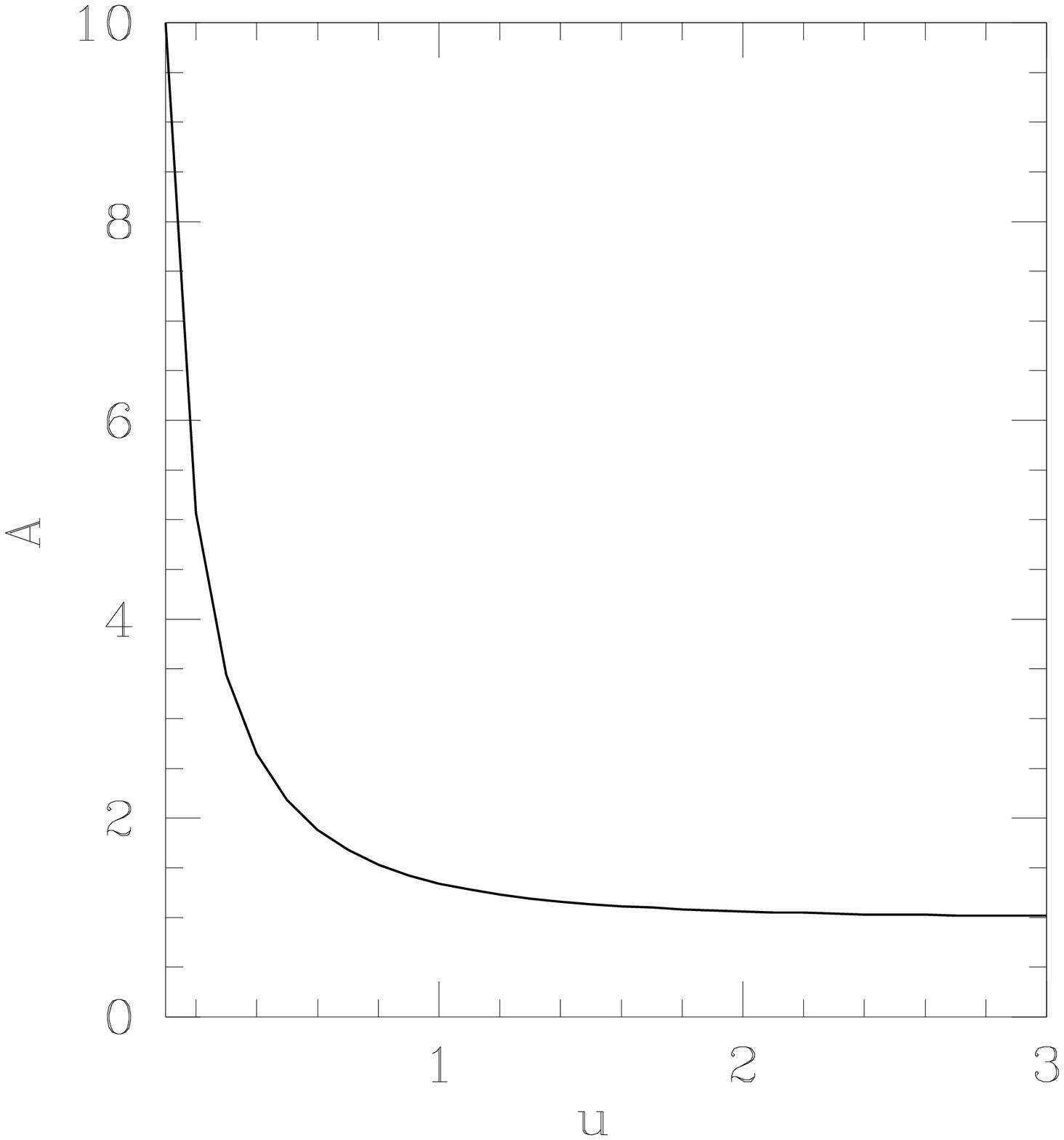}
\includegraphics[width=6.5cm]{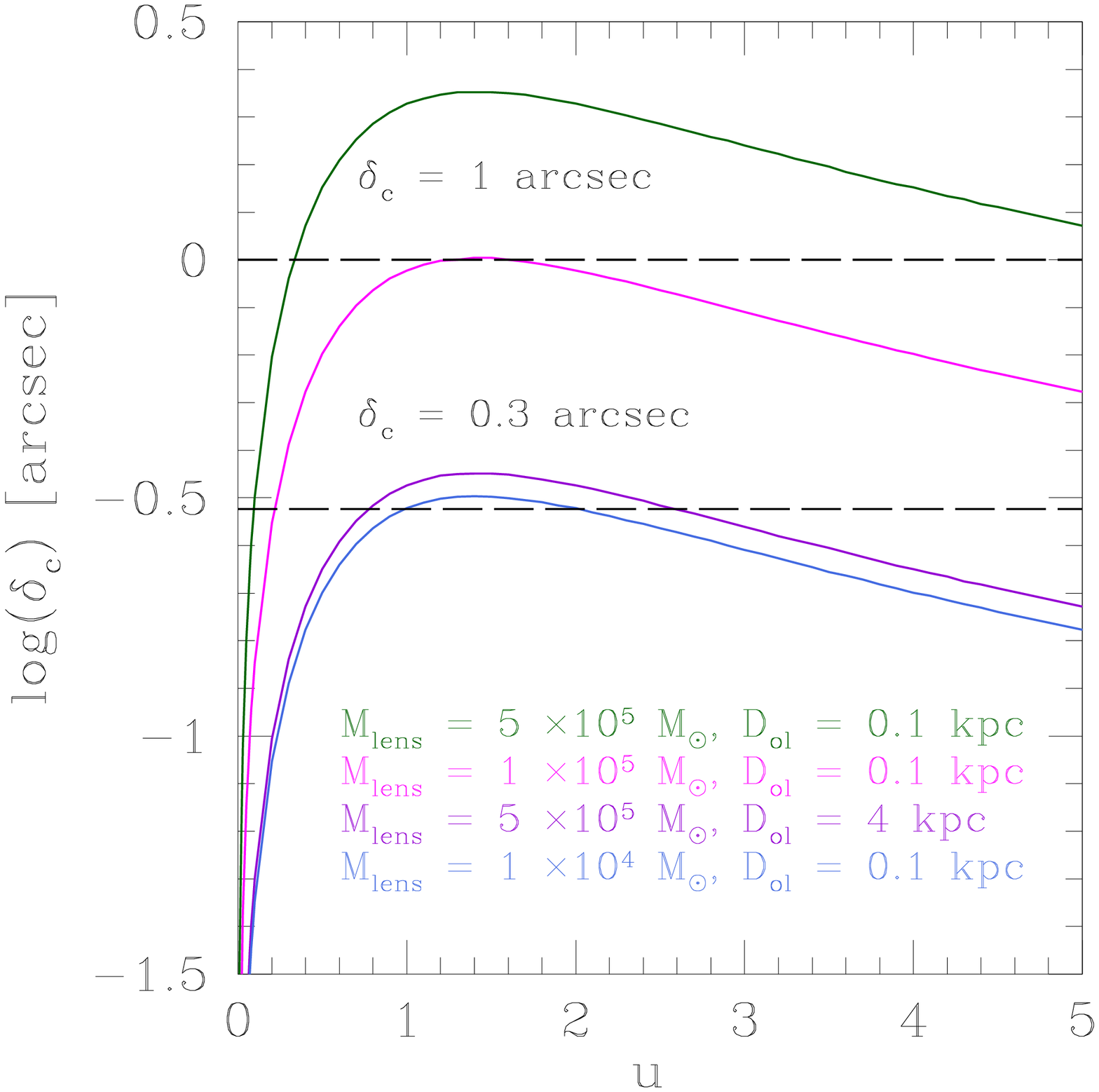}

\caption{{\it Top Left:} Distance observer-lens as a function of the Einstein radius, considering a perfectly aligned system and a point-mass lens, for five different lensing masses, and also considering a SIS density profile (grey solid line) with $\sigma_{\rm v}$ = 100 km s$^{-1}$. The vertical black solid line marks the characteristic Einstein radius $\theta_{\rm e} \approx$ 1 arcsec. {\it Top Right:} Distance observer-lens as a function of the Einstein time, for five different lensing masses. A transverse velocity of $v_{\rm T}$ = 500 km s$^{-1}$ has been assumed. The three vertical black solid lines mark the characteristic timescales $t$ = 1 and $t$ = 40 -- 60 yr. For the two {\it top} figures, distances are marked as follows: 1 AU (black dotted line), 50 AU (black dashed line; distance to the edge of the Solar System), 0.1 -- 4 kpc (black solid line; distance interval for the lens given the timescale $t$ = 1 -- 40 yr), 25 kpc (black long dashed line; distance to the edge of the Milky Way stellar disk), 150 kpc (black dot-dashed line; distance to the edge of the Milky Way halo) and 4 Mpc (black long dot-dashed line; half-way distance observer-lensed source). {\it Bottom Left:} Amplification factor as a function of the angular separation lens-lensed source, in units of the Einstein radius. {\it Bottom Right:} Astrometric shift as a function of the angular separation lens-lensed source in units of the Einstein radius, for a lensing mass of M$_{\rm lens}$ = 5 $\times$ 10$^5$ M$_{\odot}$ at a distance observer-lens of $D_{\rm ol}$ = 0.1 kpc (green solid line), M$_{\rm lens}$ = 10$^5$ M$_{\odot}$ at a distance observer-lens of $D_{\rm ol}$ = 0.1 kpc (magenta solid line), M$_{\rm lens}$ = 5 $\times$ 10$^5$ M$_{\odot}$ at a distance observer-lens of $D_{\rm ol}$ = 4 kpc (purple solid line), and M$_{\rm lens}$ = 10$^4$ M$_{\odot}$ at a distance observer-lens of $D_{\rm ol}$ = 0.1 kpc (blue solid line). The horizontal black dashed lines mark the observed proper motion/central brightness shift during 40 years ($\delta_{\rm c}$ = 0.3 arcsec and 1 arcsec).}

\end{center}
\end{figure*}


Assuming that the transient source is the result of strong lensing of a compact source (e.g., stellar cluster) within Leoncino Dwarf (Fig.~3; top), Equation~[B5] (Fig.~5; top right) can be used to constrain the distance observer-lens ($D_{\rm ol}$) based on the observed timescale ($t_{\rm e}$) of the event. An upper limit of $t \equiv t_{\rm e}  \approx$ 40 yr has been adopted, based on the timeline of the POSS images and the disappearance of the transient source, and a lower limit of $t \equiv t_{\rm e} \approx$ 1 yr has been adopted, a reasonable observational window for imaging such an event (Fig.~2; row 4 -- 5; Sect.~2.2 and 2.3). Assuming an Einstein radius of $\theta_{\rm e} \approx$ 1 arcsec (approximate distance between the transient source and the edge of the main source; Fig.~2; row 4 -- 5), and a transverse velocity of $v_{\rm T}$ = 500 km s$^{-1}$, Equation~[B5] then provides a distance observer-lens of $D_{\rm ol} \approx $ 0.1 -- 4 kpc (Fig.~5; top right); {\em the lens must be within the Milky Way halo}. Figure 5 and Equation~[B1] show that, for this distance and Einstein radius, the lens mass must be M$_{\rm lens} \approx $ 10$^4$ -- 5 $\times$ 10$^{5}$ M$_{\odot}$. Smaller lens masses at similar lens distances would imply observational windows of a month or less, while smaller lens distances would require even larger lens masses for the same timescale (Fig.~5; top left). 

If more extreme values for the Einstein radius ($\theta_{\rm e} \approx$ 10 arcsec; larger than the size of Leoncino Dwarf) and transverse velocity ($v_{\rm T}$ = 200 km s$^{-1}$) are adopted, than $D_{\rm ol} \approx $ 0.004 -- 0.2 kpc (Eq.~[B5]) and M$_{\rm lens} \approx $ 5 $\times$ 10$^4$ -- 2 $\times$ 10$^{6}$ M$_{\odot}$ (Eq.~[B1]) are obtained; these values are still within the realm of a massive lens within the Milky Way halo.

Gravitational lensing is achromatic, (de)amplifying (Eq.~[B3]) the photometric signal by the same amount regardless of the band (Fig.~5; bottom left). However, the fact that the transient source is not detectable in the red POSS I image (Fig.~2; left-hand column; row 2 -- 3) can simply reflect the bluer (rather than redder) SED\footnote{spectral energy distribution} of the lensed source. Even with amplification, it may be insufficient, in the red band, to push the image of the lensed source above the average limiting magnitude of $m_{\rm lim} \approx$ 20.0~mag of the red POSS I image (Fig.~2; left-hand column; row 2 -- 3).

Instances of (micro)lensing by small masses within the Milky Way halo are not uncommon (e.g., Alcock et al. 2001; van der Marel 2004), but these generally never exceed a few solar masses. The constrained lens mass alone (M$_{\rm lens} \approx $ 10$^4$ -- 5 $\times$ 10$^{5}$ M$_{\odot}$) excludes many Milky Way sources, such as individual stars, planets and brown dwarfs. Milky Way objects of M$_{\rm obj} \approx$ 10$^4$ -- 10$^6$ M$_{\odot}$ may, however, be stellar clusters. However, the absolute magnitudes (see also {\em Fading Stellar Cluster}) are such that they should be visible at kpc distances; for example, a globular cluster at a distance of $D \approx$ 4 kpc, would have an apparent magnitude of $m \approx$ 6 mag. Hence, stellar clusters can be discarded.

Given that the object appears to be dark, the lens may be a failed, low-mass, dark matter halo, i.e., a dark matter halo unable to retain the necessary baryons to evolve into a galaxy (e.g. Sawala et al. 2015). In this case, a NFW\footnote{Navarro-Frenk-White} (Navarro, Frenk \& White 1997), SIS (e.g., Fort \& Mellier 1994) or other density profile (e.g., Mu\~noz, Kochanek \& Keeton 2001) should be used to estimate the total lens mass. These profiles require a total lens mass much larger than that provided by the point-mass approximation (e.g., Kravtsov 2010). In this case, the required total lens mass is too large to be considered a viable possibility. Figure~5 and Equation~[B6] show that, for a SIS\footnote{singular isothermal sphere} density profile (e.g., Fort \& Mellier 1994) to produce an Einstein radius of $\theta_{\rm e} \approx$ 1 arcsec at $D_{\rm ol} \approx$ 4 kpc, requires a velocity dispersion of $\sigma_{\rm v} \approx $ 190 km s$^{-1}$. From the relation between the virial mass (M$_{\rm vir}$) and the velocity dispersion, M$_{\rm vir} \propto \sigma_{\rm v}^3$ (e.g., Munari et al. 2013), it follows that a virial mass of M$_{\rm vir} \approx$ 10$^{13}$ M$_{\odot}$ is obtained; this corresponds to a halo mass of a Milky Way-type galaxy. The application of the NFW profile provides similiar, unrealistic results. 

From this evidence, its follows that the lens must be dark, compact and massive (M$_{\rm lens} \approx$ 10$^4$ -- 5 $\times$ 10$^{5}$ M$_{\odot}$). These characteristics fall within the realm of IMBHs, in which case the lens point-mass approximation is appropriate. Within this scenario, relative motion of the lens, over a period of approximately 40 years, could induce flux variability (Sect.~2.1.4 and 2.3.1) and morphological distortions (Sect.~2.1.2; Fig.~2; row 2 -- 5) via weak lensing of the more extended parts of/other sources within Leoncino Dwarf. The brightness centroid shift ($\approx$1 arcsec; Sect.~2.1.3) provides an additional constraint if it is interpreted as a lensing astrometric shift (Eq.~[B4]) over a period of approximately 40 years. Figure~5 (bottom right) shows that astrometric shifts of $\delta_{\rm c} \approx $ 1 arcsec can be produced for $u \approx $ 1 -- 2 (angular separation lens-lensed source in units of the Einstein radius) by a M$_{\rm lens} \approx$ 10$^5$ M$_{\odot}$ lens mass at a distance of $D_{\rm ol} \approx$ 0.1 kpc. Smaller astrometric shifts can be produced for smaller distances and/or smaller lens masses (Fig.~5; bottom right).

In Section~2.4.1, a plausible explanation was provided for the behavior of the WISE data: astrometric/photometric shifts representing one weak infrared source associated with Leoncino Dwarf. In the very unlikely event that the WISE data represents true short-term variability, then a multi-component lens may be necessary. Clumpy lenses (subhalos hosted within larger halos) and secondary intergalactic low-mass field lenses have been invoked to explain anomalies in flux ratios, time delays, image separations and image distortions in quasar-galaxy lens systems (e.g., Dalal \& Kochanek 2002, Inoue \& Chiba 2005, Chen, Kravtsov \& Keeton 2003, Metcalf 2005a; Metcalf 2005b; Miranda \& Macci\`o 2007, Zackrisson \& Riehm 2010; Erickcek \& Law 2011). However, clumpy lenses and secondary lenses within the Milky Way halo would constitute highly contrived scenarios. 
 
The presence of a significant number of rogue IMBHs in galaxies in general, and in the the Milky Way bulge and halo, in particular, is theoretically predicted. First pointed out by Madau \& Rees (2001), a population of relic IMBHs naturally results from the (black hole) gravitational recoil and merging of galaxy building blocks within a hierarchical galaxy formation scenario. Subsequent numerical simulations (e.g., O'Leary \& Loeb 2009; Micic, Holley-Bockelmann \& Sigurdsson 2011; Rashkov \& Madau 2014) predict that tens to several thousand (depending on the model) IMBHs (M$_{\rm IMBH} \approx $ 10$^3$ -- 10$^6$ M$_{\odot}$) should exist in the Milky Way bulge and halo. Approximately half of these relic IMBHs are 'naked' (devoid of their halo due to tidal forces), with a more centrally concentrated (within several tens of kpc of the Galactic center) distribution. There are also predictions that such objects should be associated with old, compact ($\lesssim$1 pc), high velocity dispersion stellar cusps, but only a handful should be detectable in the visible (O'Leary \& Loeb 2012; Rashov \& Madau 2014). Results further suggest that these IMBHs may be found as (meso)lensing candidates (Micic, Holley-Bockelmann \& Sigurdsson 2011; Rashkov \& Madau 2014; Chapline \& Frampton 2016), and may be candidates for primordial black holes (Carr, K\"uhnel, \& Sandstad 2016). Oka et al. (2016, 2017) recently reported the first observational evidence for an IMBH (M$_{\rm IMBH} \approx $ 10$^5$ M$_{\odot}$) in the Milky Way using CO observations of a molecular cloud; they find a broad CO velocity width, which allows to constrain the IMBH mass, as well as a compact gas clump near the CO emission center and a point-like continuum source (see Ravi, Vedantham \& Phinney 2017, however). Tsuboi et al. (2017) report on the large velocity width of the ionized gas and compact size of the infrared IRS13E complex using ALMA\footnote{Atacama Large Millimeter Array}, which can potentially be interpreted as an ionized gas flow in Keplerian orbit around a M$_{\rm IMBH} \approx $ 10$^4$ M$_{\odot}$ IMBH. Recent work by Vedantham et al. (2017) suggest that the symmetric achromatic long-term variability in the light curves of active galaxies arises from the movement of luminal/subluminal jet features across caustics created by 10$^3$ -- 10$^6$ M$_{\odot}$ subhalo condensates or black holes within intervening galaxies.

70 -- 2000 IMBHs within the Milky Way bulge and halo (Rashov \& Madau 2014) correspond to a number density of $N \approx$ 5 $\times$ 10$^{-6}$ -- 10$^{-4}$ kpc$^{-3}$. If these IMBHs are randomly distributed within the Milky Way bulge and halo, the probability of finding one such IMBH towards a galaxy 10 arcsec in size is approximately 1 in 10$^7$; the statistical probability assigned by the current models of IMBH formation to this scenario is very low, and they would require a significant refurbishing if such a transient is produced by an IMBH.

Although the present data do not allow to firmly establish a precise interpretation for the phenomena, a lensing event appears to be the only scenario capable of simultaneously explaining the multiple changes observed in Leoncino Dwarf over a period of approximately 40 years. This includes describing the transient source as a strong lensing event, the changes in morphology and flux of the main source as time-varying weak lensing signals from the extended galaxy/compact sources within the galaxy, and the brightness centroid shift of the main source as an astrometric shift induced by a weak lensing effect. The observed parameters (timescale, flux variation, offset, size and brightness centroid shift) allow to constrain the lens system to a compact, dark lens of mass M$_{\rm lens} \approx $ 10$^4$ -- 5 $\times$ 10$^{5}$ M$_{\odot}$, likely an IMBH, at a distance observer-lens of $D_{\rm ol} \approx $ 0.1 -- 4 kpc. However, although lensing appears to explain all of the time-dependent observables (Sect.~2.1.2, 2.1.3, 2.1.4, 2.3.1 and 2.4), and notwithstanding the lensing scenario consistency with the predicted properties of relic IMBHs (mass range, distribution, compactness, potential lensing effects), the likelihood of such an event being caused by an IMBH in the Milky Way halo is still highly unlikely, and, therefore, discardable. 

\section{Conclusions}

Leoncino Dwarf is a unique source; it is an ultra-faint, extremely metal-poor dwarf galaxy, one of the current low-metallicity record-holders in the local Universe. Adding to its uniqueness, a blue transient source, to the North of the main source, appears to be present in images from 1955. The main source appears also to show a change in morphology, a brightness centroid shift, and peak flux variability ($\Delta m \simeq$ 0.7 mag) over a period of approximately 40 years. Variability of massive stars resolved in the HST images could, in principle, explain the brightness centroid shift, morphology change and (long- and short-term) peak flux variability of the main source, but this scenario requires an independent explanation for the transient. The passing of a Solar System object, a fading stellar cluster, dust enshroudment by feedback processes, and variable accretion processes onto a compact object have been investigated, and excluded, as possible explanations for the transient phenomena. A lensing event could provide an interesting alternative, simultaneous explanation for all the observables. The timescale for the transient and other variable observables requires a dark, compact, massive (M$_{\rm lens} \approx $ 10$^4$ -- 5 $\times$ 10$^{5}$ M$_{\odot}$) lens, likely a IMBH, within the Milky Way halo ($D_{\rm ol} \approx $ 0.1 -- 4 kpc). Strong lensing describes the transient source, while weak lensing of the extended galaxy/compact sources within the galaxy describes the morphological and peak flux changes, as well as the observed brightness centroid shift (astrometric shift). However, as this scenario is statistically highly unlikely according to current theoretical predictions, it can also be discarded. The transient could have been the result of a cataclysmic event such as a SN or hypernova, as long as the event was caught in the later/late stages of the light curve. However, in these cases, a (net) magnitude variation of $\Delta m \approx$ 1 -- 2 mag of the transient over an approximately 60-year period should have produced a detectable quiescent source and/or the host HII region should have deen detected in the SDSS and HST images. An episode related to a stellar merger/nova or giant eruption of a LBV star, observed at, or near, peak magnitude, may explain the transient source, and, possibly, the brightness centroid shift/morphology change, but can not account for the remaining observable, the peak flux variation of the main source, unless massive star variability in the main source is evoked.
 
\section*{Acknowledgments}

The authors would like to thank Evencio Mediavilla for useful discussions on gravitational lensing, Joana Ascenso for help on the PSF modeling, Santiago Gonz\'alez-Gait\'an for pointing out calcium-rich SN, and Dave Monet for valuable guidance regarding the treatment of POSS data. The authors would also like to thank the anonymous reviewer for comments and suggestions that have greatly improved this manuscript. M. E. F. gratefully acknowledges the financial support of the ''Funda\c c\~ao para a Ci\^encia e Tecnologia'' (FCT -- Portugal), through the research grant SFRH/BPD/107801/2015. This work has been partly funded by the Spanish Ministery of Economy and Competitiveness, project {\em Estallidos} AYA2013-47742-C04-02-P, AYA2013-47742-C04-01-P and AYA2016-79724-C04-2-P. This work has made use of the WISE, 2MASS, FIRST, NVSS, VLSS, Herschel, Chandra and XMM-Newton databases, the NASA Jet Propulsion Laboratory's HORIZONS Web-Interface, the SDSS DR12, STScI digitized POSS data, the NASA/IPAC Extragalactic Database (NED), VizieR and the Ned Wright's Cosmological Calculator. We have also consulted the IAU Central Bureau for Astronomical Telegrams.

\section*{References}


\noindent  Alcock, C., Allsman, R. A., Alves, D. R. et al. 2001, Nature, 414, 617 

\noindent  Allen, L., Megeath, S. T., Gutermuth, R., Myers, P. C., Wolk, S., Adams, F. C., Muzerolle, J., Young, E. \& Pipher, J. L. 2007, Protostars and Planets V, B. Reipurth, D. Jewitt \& K. Keil (eds.), University of Arizona Press, Tucson, p. 361 -- 376

\noindent  Alvarez, M. A., Bromm, V. \& Shapiro, P.R. 2006, ApJ, 639, 621

\noindent  Annunziatella, M., Mercurio, A., Brescia, M., Cavuoti, S. \& Longo, G. 2013, PASP, 125, 68

\noindent  Arcavi, I., Howell, D. A., Kasen, D., Bildsten, L., Hosseinzadeh, G. et al. 2017, Nature, 551, 210

\noindent  Bassa, C. G., Tendulkar, S. P., Adams, E. A. K. et al. 2017, ApJL, 843, 8

\noindent  Bennett, C. L., Larson, D., Weiland, J. L. \& Hinshaw, G. 2014, ApJ, 794, 135

\noindent  Bertin, E. \& Arnouts, S. 1996, A\&AS, 117, 393

\noindent  Bogdanovi\'c, T., Eracleous, M. Mahadevan, S., Sigurdsson, S., \& Laguna, P. 2004, ApJ, 610, 707

\noindent  Bomans, D. J. \& Weis, K. 2011, Bulletin Soci\'et\'e Royale des Sciences de Li\'ege, Proceedings of the 39th Li\'ege Astrophysical Colloquium,  G. Rauw, M. De Becker, Y. Naz\'e, J. -M. Vreux \& P. Williams (eds.), vol. 80, p. 341-345

\noindent  Brodie, J. P. \& Strader, J. 2006, ARA\&A, 44, 193

\noindent  Brunthaler, A., Reid, M. J., Falcke, H., Henkel, C. \& Menten, K. M. 2007, A\&A, 462, 201

\noindent  Carr, B., K\"uhnel, F. \& Sandstad, M. 2016 2016, PhRvD, vol. 94, issue 8

\noindent  Casares, J., Jonker, P. G. \& Israelian, G. 2017, Handbook of Supernovae, ISBN 978-3-319-21845-8, Springer International Publishing AG, p. 1499


\noindent  Chambers, K. C., Magnier, E. A., Metcalfe, N., Flewelling, H. A., Huber, M. E. et al. 2016, arXiv:1612.05560

\noindent  Chapline, G. \& Frampton, P. H. 2016, JCAP, 11, 42

\noindent  Chatterjee, S., Law, C. J., wharton, R. S. et ak. 2017, Nature, 541, 58

\noindent  Chen, J., Kravtsov, A. V. \& Keeton, C. R. 2003, ApJ, 592, 24

\noindent  Cheung, E., Bundy,	K. Cappellari, M. et al. 2016, Nature,  533, 504

\noindent  Childress, M., Aldering, G., Antilogus, P., Aragon, C., Bailey, S. et al. 2013, ApJ, 770, 107

\noindent  Dalal, N. \& Kochanek, C. S. 2002, ApJ, 572, 25

\noindent  di Matteo, T., Springel, V., Hernquist, L. et al. 2005, Nature, 433, 604

\noindent  Djorgovski, S. G., Drake, A. J., Mahabal, A. A., Graham, M. J., Donalek, C. et al. 2011, arXiv:1102.5004

\noindent  Djorgovski, S. G., Mahabal, A. A., Drake, A. J., Graham, M. J., Donalek, C. \& Williams, R. 2012, IAUS, 285, 141

\noindent  Djorgovski, S. G., Mahabal, A., Drake, A., Graham, M. \& Donalek, C. 2013, Planets, Stars and Stellar Systems, by Oswalt, Bond \& Howard, ISBN 978-94-007-5617-5. Springer Science+Business Media Dordrecht, p. 223

\noindent  Dugan, Z., Bryan, S., Gaibler, V., Silk, J., Haas, M. 2014, ApJ, 796, 113

\noindent  Erickcek, A. L. \& Law, N. M. 2011, ApJ, 729, 49

\noindent  Filho, M. E., Winkel, B., S\'anchez Almeida, J. et al. 2013, A\&A, 558, 18

\noindent  Filho, M. E., S\'anchez Almeida, J., Mu\~noz-Tu\~n\'on, C., Nuza, S. E., Kitaura, F. \& He\ss, S. 2015, ApJ, 802, 82

\noindent  Filho, M. E., S\'anchez Almeida, J., Amor\'\i n, R.,  Mu\~noz-Tu\~n\'on, C., Elmegreen, B. G. \& Elmegreen, D. M. 2016, ApJ, 820, 109

\noindent  Firth, R. E., Sullivan, M., Gal-Yam, A. et al. 2015, MNRAS, 446, 3895


\noindent  Foley, R. J., Chornock, R., Filippenko, A. V., Ganeshalingam, M., Kirshner, R. P., 2009, AJ, 138, 376

\noindent  Foley, R. J. 2015, MNRAS, 452, 2463

\noindent  Fort, B. \& Mellier, Y. 1004, A\&ARv, 5, 239

\noindent  Fruchter, A. S., Levan, A. J., Strolger, L. et al. 2006, Nature, 441, 463


\noindent  Gabor, J. M. \& Bournaud, F. 2013, MNRAS, 434, 606


\noindent  Gould, A. 2000, ApJ, 542, 785

\noindent  Groh, J. H., Meynet, G., Ekstr\"{o}m, S. \& Georgy, C. 2014, A\&A, 564, 30

\noindent  Hachisu, I. \& Kato, M. 2015, ApJ, 798, 76

\noindent  Hamuy, M., Folatelli, G., Morrell, N. I., Phillips, M. M., Suntzeff, N. B. et al. 2006, PASP, 118, 2 

\noindent  Hawkins, M. R. S. 1996, MNRAS, 278, 787

\noindent  Hayashi, E. \& White, S. D. M. 2008, MNRAS, 388, 2

\noindent  Hirschauer, A. S., Salzer, J. J., Skillman, E. D., Berg, D., McQuinn, K. B. W. et al. 2016, ApJ, 822, 108

\noindent  Hodgkin, S. T., Wyrzykowski, L., Blagorodnova, N. \& Koposov, S. 2013, Philosophical Transactions of the Royal Society A: Mathematical, Physical and Engineering Sciences, vol. 371, issue 1992, p. 20120239-20120239

\noindent  Hopkins, P. F., Henrquist, L., Cox, T. J. et al. 2005, ApJ, 630, 705

\noindent  Inoue, K. T. \& Chiba, M. 2005, ApJ, 634, 77


\noindent  Ivezi\' c, Z., Tyson, J. A., Abel, B., Acosta, E.  et al. 2008, arXiv:0805.2366


\noindent  Izotov, Y. I. \& Thuan, T. X. 2008, ApJ, 687, 133

\noindent  Izotov, Y. I. \& Thuan, T. X. 2009, 690, 1797

\noindent  Izotov, Y. I., Thuan, T. X. \& Guseva, N. G. 2007, ApJ, 671, 1297

\noindent  Izotov, Y. I., Thuan, T. X., Guseva, N. G. \&  Liss, S. E. 2018, MNRAS, 473, 1956



\noindent  Juri\'c, M., Ivezi\'c, \v{Z}. Lupton, R. H., Quinn, T. Tabachnik, S. 2002, AJ, 124, 1776 

\noindent  Kasen, D. \& Woosley, S. E. 2009, ApJ, 703, 2205

\noindent  Kokubo, M., Mitsuda, K., Sugai, H., Ozaki, S., Minowa, Y. et al. 2017, ApJ, 844, 95

\noindent  Kravtsov, A. 2010, AdAst2010

\noindent  Kochanek, C. S., Szczygie\l, D. M. \& Stanek, K. Z. 2012, ApJ, 758, 142

\noindent  Kochanek, C. S., Adams, S. M. \& Belczynski, K. 2014, MNRAS, 443, 1319

\noindent  Lasker, B. M., Lattanzi, M. G., McLean, B. J. et al. 2008, AJ, 136, 735

\noindent  Law, N. M., Kulkarni, S. R., Dekany, R. G., Ofek, E. O., Quimby, R. M. et al. 2009, PASP, 121, 1395

\noindent  Lorimer, D. R., Bailes, M., McLaughlin, M. A., Narkevic, D. J. \& Crawford, F. 2007, Science, 318, 777 

\noindent  Leloudas, G., Schulze, S., Kr\"{u}hler, T., Gorosabel, J., Christensen, L. et al. 2015, MNRAS, 449, 917

\noindent  Lunnan, R., Chornock, R., Berger, E. et al. 2014, ApJ, 787, 138

\noindent  Lunnan, R., Chornock, R., Berger, E. et al. 2015, ApJ, 804, 90 

\noindent  Lunnan, R., Kasliwal, M. M., Cao, Y. et al. 2017, ApJ, 836, 60

\noindent  Lyman, J. D., Levan, A. J., James, P. A., Angus, C. R., Church, R. P., Davies, M. B. \& Tanvir, N. R. 2016, MNRAS, 458, 1768

\noindent  Lyman, J. D., Levan, A. J., Tanvir, N. R., Fynbo, J. P. U., McGuire, J. T. W. et al. 2017, MNRAS, 467, 1795

\noindent  Madau, P. \& Rees, M. J. 2001, ApJ, 551, 27

\noindent  Maeda, K., Mazzali, P. A., Deng, J. et al. 2003, ApJ, 593, 931

\noindent  Marcote, B., Paragi, Z., Hessels, J. W. T. et al. 2017, ApJL, 834, 8

\noindent  Massey, P. 2010, ASPC, 425, 3

\noindent  Metcalf, R. B. 2005a, ApJ, 629, 673

\noindent  Metcalf, R. B. 2005b, ApJ, 622, 72

\noindent  McHardy, I. M., Connolly, S. D., Peterson, B. M. et al. 2016, AN, 337, 500

\noindent  Micic, M., Holley-Bockelmann, K. \& Sigurdsson, S. 2011, MNRAs, 414, 1127

\noindent  Miranda, M. \& Macci\`o, A. V. 2007, MNRAS, 382, 1225

\noindent  Maoz, D., Mannucci, F. \& Nelemans, G. 2014, ARA\&A, 52, 107

\noindent  Modjaz, M., Kewley, L., Krischner, R. P., Stanek, K. Z, Challis, P. et al. 2008, AJ, 135, 1136 

\noindent  Monet, D. G. 1998, AAS, 193rd AAS Meeting, id.120.03; Bulletin of the American Astronomical Society, Vol. 30, p.1427

\noindent  Monet, D. G., Levine, S. E., Mayzian, B. et al. 2003, AJ, 125, 984

\noindent  Mu\~noz, J. A., Kochanek, C. S. \& Keeton, C. R. 2001, ApJ, 558, 657

\noindent  Munari, E., Biviano, A., Borgani, S., Murante, G. \& Fabjan, D. 2013, MNRAS, 430, 2638

\noindent  Naz\'e, Y., Rauw, G. \& Hutsem\'ekers, D. 2012, A\&A, 538, 47 

\noindent  Navarro, J. F., Frenk, C. S. \& White, S. D. 1997, ApJ, 490, 493

\noindent  Nayakshin, S. 2014, MNRAS, 437, 2004

\noindent  Nierenberg, A. M., Treu, T., Menci, N., Lu, Y., Torrey, P. Vogelsberger, M. 2016, MNRAS, 462, 4473

\noindent  Nomoto K., Maeda K., Mazzali P.A., Umeda H., Deng J., Iwamoto K. 2004, Hypernovae and Other Black-Hole-Forming Supernovae, Fryer C. L. (eds.) Stellar Collapse, Astrophysics and Space Science Library, vol 302. Springer, Dordrecht

\noindent  Novak, G. S., Ostriker, J. P. \& Ciotti, L. 2011, APJ, 737, 26

\noindent  O'Leary, R. M. \& Loeb, A. 2012, MNRAS, 395, 781

\noindent  O'Leary, R. M. \& Loeb, A. 2012, MNRAS, 421, 2737

\noindent  Olmo-Garc\'\i a, A., S\'anchez Almeida, J., Mu\~noz-Tu\~n\'on, C., Filho, M. E., Elmegreen, B. G., Elmegreen, D. M., P\'erez-Montero, E. \& M\'endez-Abreu, J. 2017, ApJ, 834, 181

\noindent Oka, T., Mizuno, R., Miura, K. \& Takekawa, S., ApJ, 816, L7

\noindent Oka, T., Tsujimoto, S., Iwata, Y., Nomura, M. \& Takekawa, S. 2017, Nature Astronomy, 1, 709

\noindent  Peacock, M. B., Maccarone, T. J.m Kundi, A. \& Zepf, S. 2010, MNRAS, 407, 2611


\noindent  Pustilnik, S. A., Tepliakova, A. L., Kniazev, A. Y. \& Burenkov, A. N. 2008, MNRAS, 388, 24

\noindent  Pustilnik, S. A., Makarova, L. N., Perepelitsyna, Y. A., Moiseev, A. V. \& Makarov, D. I. 2017, MNRAS, 465, 4985

\noindent  Quimby, R. M., Kulkarni, S. R., Kasliwal, M. M., Gal-Yam, A., Arcavi, I. et al. 2011, Nature, 474, 487

\noindent  Rashkov, V. \& Madau, P. 2014, ApJ, 780, 187

\noindent  Rau, A., Kulkarni, S. R., Law, N. M., Bloom, J. S., Ciardi, D. et al. 2009, PASP, 121, 1334

\noindent  Ravi, V., Vedantham, H. \& Phinney, E. S. 2017, arXiv:1710.03813 

\noindent  Read, J. I., Iorio, G., Agertz, O., \& Fraternali, F. 2016, MNRAS, 462, 3628

\noindent  Reid, I. N., Brewer, C., Brucato, R. J., McKinley, W. R., Maury, A. et al. 1991, PASP, 103, 661 

\noindent  Reines, A. E., Greenem J. E. \& Geha, M. 2013, ApJ, 775, 116

\noindent  Roeser, S., Demleitner, M. \& Schilbach, E. 2010, AJ, 139, 2440

\noindent  Rogerson, J. A., Hall, P. B., Hildago, P. R. et al. 2015, MNRAS, 457, 405

\noindent  Shafter, A. W., Curtin, C., Pritchet, C. J., Bode, M. F. \& Darnley, M. J. 2014, ASPC, 490, 77

\noindent  S\'anchez Almeida, J., Filho, M. E., Dalla Vecchia, C. \& Skillman, E. D. 2017, ApJ, 835, 159

\noindent  Sawala, T., Frenk, C. S., Fattahi, A. et al. 2015, MNRAS, 448, 2941

\noindent  Scalzo, R., Yuan, F., Childress, M. J. et al. 2017, PASA, 34, 30


\noindent  Siemiginowska, A. \& Elvis, M. 1997, ApJ, 482, 9


\noindent  Smartt, S. J. 2009, ARA\&A, 47, 63 

\noindent  Smith, A. M., Lynn, S., Sullivan, M., Lintott, C. J., Nugent, P. E. et al. 2011a, MNRAS, 412, 1309 

\noindent  Smith, N., Li, W., Silverman, J. M., Ganeshalingam, M. \& Filippenko, A. V. 2011b, MNRAS, 415, 773 

\noindent  Smith, N., Andrews, J. E., Van Dyk, S. D. et al. 2016, MNRAS, 458, 950

\noindent  Smith, N. 2016, MNRAS, 461, 3353

\noindent  Smith, N \& Tombleson, R. 2015, MNRAs, 447, 598

\noindent  Sterken, C. 2003, ASPC, 292, 437

\noindent  Sutton, A. D., Roberts, T. P., Walton, D. J., Gladstone, J. C. \& Scott, A. E. 2012, MNRAS, 423, 1154

\noindent  Tang, S., Bildsten, L., Wolf, W. M. et al. 2014, ApJ, 786, 61

\noindent  Tendulkar, S. P., Bassa, C. G., Cordes, J. M. et al. 2017, ApJL, 834, 7

\noindent  Tsuboi, M., Kitamura, Y., Tsutsumi, T., Uehara, K., Miyoshi, M., Miyawaki, R. \& Miyazaki, A. 2017, ApJ, 850, 5


\noindent  van der Marel, R. P., Anderson, J., Bellini, A. et al. 2014, ASPC, 480, 43

\noindent  van der Marel, R. P., Coevolution of Black Holes and Galaxies, from the Carnegie Observatories Centennial Symposia. Published by Cambridge University Press, as part of the Carnegie Observatories Astrophysics Series. Edited by L. C. Ho, 2004, p. 37

\noindent  Vedantham, H. K., Readhead, A. C. S., Hovatta, T. et al. 2017, 845, 89

\noindent  Walborn, N. R., Howarth, I. D., Lennon, D. J. et al. 2002, ApJ, 123, 2754 

\noindent  Zackrisson, E. \& Riehm, T. 2010, AdAst2010

\noindent  Zuo, W., Wu, X.-B., Liu, Y.-Q., Jiao, C.-L. 2012, ApJ, 758, 104


\appendix

\section{Transient Flux Error Estimation}

The integrated signal of the transient, $S$, is approximately given by

\begin{equation}
S\simeq n_{pix} \, f\ I_{max},
\end{equation}

\noindent where $I_{max}$, $n_{pix}$ and $f$ represent the maximum signal, the number of pixels encompassed by the source, and a factor of the order unity, respectively; the latter accounts for the variation of the signal in the pixels surrounding the maximum. On the other hand, the noise on the integrated signal, $\Delta S$, can be expressed as

\begin{equation}
\Delta S \simeq \sqrt{n_{pix}} \,\Delta I,
\label{eq:this}
\end{equation}  
 
\noindent where $\Delta I$ represents the error in the intensity in a single pixel. Equation ([A1]) assumes that the noise in the different pixels contributing to $S$ is similar and independent. Combining the two expressions above, the relative error of the integrated flux is

\begin{equation}
\frac{\Delta S}{S}\simeq \frac{1}{f\,\sqrt{n_{pix}}}\,\frac{\Delta I}{I_{max}}.
\label{eq:that}
\end{equation}

If the transient has a FWHM of approximately 4~arcsec, and is detected with a pixel size of $\sim$1 arcsec (Sect.~2.1.1), then $n_{pix}\simeq 16$. A value of $\Delta I/I_{max}\simeq 1/5$ is measured (Sect.~2.1.2) which, together with Equation~([A3]), leads to $\Delta S / S$ values between 0.05 and 0.10 when $f$ varies between 1 and 0.5, which is the range of signals within the FWHM of the source. Because the above estimate is only approximate, a conservative limit of

\begin{equation}
\frac{\Delta S}{S}\simeq 0.2
\end{equation} 

\noindent has been adopted (Sect.~2.1.2).

\section{Gravitational Lensing Equations}

The lensing of a source requires a set of specific circumstances in order to occur, namely, the relative alignment between the observer, the lens and the lensed source. Strong lensing effects, such as high magnifications, strong image distortions and mutiple images, are produced when the lens is well-aligned with the line-of-sight. If the lensed source is extended, strong lensing will produce multiple arcs and rings, while if the lensed source is point-like, such as a quasar, it will produce point-like images. Weak lensing occurs when the lens is farther from the line-of-sight. In this case, a single image is produced, subject to mild magnification and distortion. 

When the lensed source, center of the lens and the observer are perfectly aligned, the lensed source appears to the observer as a ring. The radius of the ring, the Einstein radius ($\theta_{\rm e}$), provides a characteristic length-scale for the lensing phenomenon. The lensing equation (e.g., Hawkins 1996; Gould 2000) provides a relation between the Einstein radius, the lens mass (M$_{\rm lens}$), and the relative angular distances ($R$). For a point-mass lens

\begin{equation}
\theta_{\rm e} = \Big( \frac{4 \, G \, {\rm M_{lens}}}{c^2 \, R} \Big)^{\frac{1}{2}},
\label{eq:lens}
\end{equation}

\noindent where $G$ is the gravitational constant, $c$ is the speed of light, and

\begin{equation}
R = \frac{D_{\rm ol} \, D_{\rm os}}{D_{\rm ls}},
\end{equation}

\noindent where $D_{\rm ol}$ is the distance observer-lens, $D_{\rm os}$ is the distance observer-lensed source and $D_{\rm ls}$ is the distance lens-lensed source. For low-redshift sources, $D_{\rm ls} \simeq D_{\rm os} - D_{\rm ol}$. If the lensed source is misaligned from the observer-lens axis by some angle $\beta$, which may be a function of time ($t$), the total photometric amplification ($A$) of the lensed source due to a point-mass lens is

\begin{equation}
A = \frac{u^2 + 2}{u \sqrt{u^2 + 4}},
\end{equation}

\noindent where $u \equiv u(t) = \frac{\beta}{\theta_{\rm e}}$ is the angular separation lens-lensed source in Einstein radius units. The center of brightness is also shifted ($\delta_{\rm c}$) relative to the true position of the lensed source by the point-mass lens 

\begin{equation}
\delta_{\rm c} = \theta_{\rm e} \frac{u}{u^2 + 2}.
\end{equation}

\noindent For a lens moving across the line-of-sight with a transverse velocity ($v_{\rm T}$), the timescale ($t_{\rm e}$) for the variation (Einstein time), i.e., the time to cross the Einstein radius, is given by

\begin{equation}
t_{\rm e} = \frac{D_{\rm ol} \, \theta_{\rm e}}{v_{\rm T}}.
\end{equation}

\noindent If, instead of a point mass, a SIS (e.g., Fort \& Mellier 1994) density profile is considered for the lens, a profile commonly explored in lensing studies, then the lensing equation becomes

\begin{equation}
\theta_{\rm e} = \frac{4 \, \pi \, \sigma_{\rm v}^2}{c^2} \, \frac{D_{ls}}{D_{os}},
\end{equation} 

\noindent where $\sigma_{\rm v}$ is the velocity dispersion. The lens Equation ([B1]) still holds for an extended lens, but, in this case, the lens mass corresponds only to the mass contained within the Einstein radius and not the total lens mass. 


\bsp	
\label{lastpage}
\end{document}